%% file: Main.tex
\documentclass[aps,pra,superscriptaddress, preprint,floatfix,longbibliography,showpacs,amsfonts,amsmath,amssymb]{revtex4-1}
\usepackage{graphicx}
\usepackage{hyperref}
\usepackage{xr-hyper}
\usepackage{upgreek}
\usepackage{epstopdf}
\usepackage[note-name=, use-sort-key = false]{notes2bib}

\begin{document}

\title{Easy-cone state mediating the spin reorientation in topological kagome magnet Fe$_3$Sn$_2$}

\author{L. Prodan}
\affiliation{Experimentalphysik V, Center for Electronic Correlations and Magnetism, Institute of Physics, University of Augsburg, D-86159 Augsburg, Germany}
\author{ D. M. Evans}
\affiliation{Experimentalphysik V, Center for Electronic Correlations and Magnetism, Institute of Physics, University of Augsburg, D-86159 Augsburg, Germany}
\affiliation{Department of Sustainable Energy Technology, SINTEF Industry, Oslo, Norway}
\author{ A. S. Sukhanov}
\affiliation{Experimentalphysik VI, Center for Electronic Correlations and Magnetism, Institute of Physics, University of Augsburg, D-86159 Augsburg, Germany}
\author{S.~E.~Nikitin}
\affiliation{PSI Center for Neutron and Muon Sciences, Paul Scherrer Institut, CH-5232 Villigen-PSI, Switzerland}
\author{A. A. Tsirlin}
\affiliation{Felix Bloch Institute for Solid-State Physics, Leipzig University, D-04103 Leipzig, Germany}
\author{ L. Puntingam}
\affiliation{Experimentalphysik V, Center for Electronic Correlations and Magnetism, Institute of Physics, University of Augsburg, D-86159 Augsburg, Germany}
\author{M. C. Rahn}
\affiliation{Experimentalphysik VI, Center for Electronic Correlations and Magnetism, Institute of Physics, University of Augsburg, D-86159 Augsburg, Germany}
\author{L. Chioncel}
\affiliation{Theoretische Physik III,  Institute of Physics, University of Augsburg, D-86135 Augsburg, Germany}
\author{V. Tsurkan}
\affiliation{Experimentalphysik V, Center for Electronic Correlations and Magnetism, Institute of Physics, University of Augsburg, D-86159 Augsburg, Germany}
\affiliation {Institute of Applied Physics, Moldova State University, MD 2028 Chișinău, R. Moldova}

\author{I. K\'ezsm\'arki}
\affiliation {Experimentalphysik V, Center for Electronic Correlations and Magnetism, Institute of Physics, University of Augsburg, D-86159 Augsburg, Germany}

\begin{abstract}
We investigated temperature-driven spin reorientation (SR) in the itinerant kagome magnet Fe$_3$Sn$_2$ using high-resolution synchrotron x-ray diffraction, neutron diffraction, magnetometry, and magnetic force microscopy (MFM), further supported by phenomenological analysis. Our study reveals a crossover from the state with easy-plane anisotropy to the high-temperature state with uniaxial easy-axis anisotropy taking place between $\sim40-130$~ K through an intermediate easy-cone (or tilted spin) state. This state, induced by the interplay between the anisotropy constants $K_1$ and $K_2$, is clearly manifested in the thermal evolution of the magnetic structure factor, which reveals a gradual change of the SR angle $\mathbf{\theta}$ between $40-130$~K. We also found that the SR is accompanied by a magnetoelastic effect. Zero-field MFM images across the SR range show a transformation in surface magnetic patterns from a dendritic structure at 120~K, to domain wall dominated MFM contrast at 40~K.

*To whom correspondence should be addressed. Email: lilian.prodan@uni-a.de
\end{abstract}


\maketitle
\section{Introduction}

In anisotropic magnetic materials, the easy axis of magnetization denotes the direction along which magnetic moments align, minimizing the internal energy of the system~\cite{Buschow2003}. The interplay between magnetocrystalline anisotropy, dipolar, and Dzyaloshinskii-Moriya interactions (DMI) might lead to a reorientation of magnetic moments~\cite{Dzyaloshinsky1958, Moriya1960}. The possibility of controlling magnetic degrees of freedom through the interplay between different types of interactions has been the focus of extensive studies in recent decades ~\cite{Locatelli2013, Zuo2021, Kisi2024, Pal2024}. This has been extensively studied for thin films and multilayered structures where long-range magnetic dipolar interaction dominates, forcing magnetic moments to follow shape anisotropy~\cite{Johnson1996}. On the other hand, in RE-TM magnets (RE~ =~ Pr, Nd, Tb, Dy, Ho; TM~ =~ Mn, Fe, Co) the competition between the anisotropies of rare-earth and transition-metal sublattices drives the so-called spin reorientation (SR) transition~\cite{White1969, Yamaguchi1974, Algarabel1988, Huang2024}, where the direction of the easy axis of magnetization rotates from one crystallographic axis to another with varying the temperature. Moreover, temperature-induced changes in the local environment and magnetic interactions between ions at different Wyckoff positions due to symmetry change can also induce spin reorientation effects~\cite{Weber2022}.

Depending on whether the easy axis shifts abruptly or rotates continuously between crystal symmetry axes, the reorientation is characterized as a first- or second-order phase transition, respectively, or can even be a smooth crossover ~\cite{Levinson1969}. A first-order magnetic transition is usually accompanied by a hysteresis in the magnetization and substantial changes in the magnetic entropy~\cite{Belov1976, Yuan2013, Song2024}. In contrast, second-order phase transitions can be traced as distinct anomalies in the magnetic susceptibility, while crossovers may not be accompanied by any sharp anomaly in thermodynamic quantities ~\cite{Belov1976, Law2018, Horner1968}.  

Recently, there has been growing interest in materials exhibiting SR, where the orientation of the easy axis can be influenced by a range of intrinsic and extrinsic factors, including chemical composition ~\cite{Moore2022, Karube2022}, shape~\cite{Johnson1996}, magnetic field ~\cite{Masuda2020, Yamamoto2021, Prodan2021}, external pressure~\cite{Lin2018, Skorobogatov2023}, strain~\cite{Bordel2012, Kong2024}, and light ~\cite{Kimel2004}. These factors often involve significant changes in the magnetic and electronic properties of materials and lead to the emergence of novel quantum phenomena. In particular, in several non-centrosymmetric magnets, it has been demonstrated that the interplay between DMI and magnetic anisotropy is crucial for stabilizing topologically protected (anti)skyrmion states~\cite{Kezsmarki2015, Ehlers2016, Bordacs2017, Leonov2017, Preissinger2021,  Zuo2021, Meshcheriakova2014, Sukhanov2020, Karube2022}. Lorentz force microscopy studies on Nd$_2$Fe$_{14}$B have revealed the formation of magnetic bubbles with zero topological number and magnetic skyrmions induced by SR~\cite{Xiao2020}. Moreover, the enhancement of magnetoresistance, topological, and anomalous Hall effects have been reported near the SR transition in various two-dimensional van der Waals magnets~\cite{Pal2024, Bera2023, Wang2024},  rare-earth topological kagome magnets of RMn$_6$Sn$_6$-type ~\cite{Jones2024, Lv2025}, and Heusler magnets ~\cite{Kumar2020}. These phenomena are attributed to changes in Berry curvature during spin reorientation ~\cite{He2020}.

Here, we investigate the SR effect in the itinerant kagome magnet Fe$_3$Sn$_2$. The crystal structure consists of bilayers of Fe$_3$Sn kagome networks separated by hexagonal Sn layers and stacked along the \textbf{\textit{c}} axis [see Fig.~\ref{fig:fig1XRD}(a)], corresponding to rhombohedral $R\overline{3}m$ space group~\cite{Malaman1976}. Fe$_3$Sn$_2$ is ferromagnetic below 650~K and its magnetic anisotropy is reported to change from the easy-axis (EA) above 250~K to the easy-plane (EP) below 60~K~\cite{Malaman1978, He2021}. This material has recently attracted much attention as a host for massive Dirac fermionic states ~\cite{Ye2018}, Weyl nodes~\cite{Biswas2020}, topological flat bands~\cite{Lin2018a}, and helical nodal lines ~\cite{Schilberth2022}. The interplay between the electronic band topology and magnetism gives rise to prominent phenomena, including large anomalous and topological Hall effects~\cite{Ye2018, Du2022}, magneto-optical effects ~\cite{Biswas2020, Schilberth2022}, and topologically protected skyrmionic bubbles~\cite{Hou2018, Altthaler2021}. Moreover, recent advancements in the field suggested that Fe$_3$Sn$_2$ manifests anomalous charge carrier behavior and strong orbital contribution to the magnetism and transport properties~\cite{Ekahana2024, GoncalvesFaria2024, Zhang2024, Wang2024_2}.

Despite intense studies using various measurement techniques such as magnetization, \textit{ac}-susceptibility, neutron diffraction, magnetic force microscopy (MFM) ~\cite{Fenner2009, Fayyazi2019, Heritage2020, Wu2021, Xie2024}, optical spectroscopy~\cite{Biswas2020}, Raman scattering ~\cite{Wu2021, He2021}, and magnetotransport measurements~\cite{Wang2016, Kumar2019}, the mechanism of spin reorientation in Fe$_3$Sn$_2$ remains unclear. Specifically, whether the transition is gradual or abrupt, its origin, and its impact on the magnetic domain configuration are still debated. Given that the electronic band topology can strongly be affected by the magnetic structure in kagome magnets, like Fe$_3$Sn$_2$, it is essential to clarify the origin of the SR and understand the accompanying variations in spin texture with temperature and magnetic field.

In this study, we investigated the temperature-driven SR in the topological kagome ferromagnet Fe$_3$Sn$_2$ through a combination of experimental techniques and phenomenological analysis. The temperature dependence of the anisotropy constants $K_1$ and $K_2$, the evolution of the magnetic structure factor $\mathbf{M_Q}(T)$, and MFM imaging indicate that the high-temperature state, with uniaxial easy-axis anisotropy, transforms to a mixed state before developing an in-plane magnetization state at low temperatures, consistent with easy-plane anisotropy. This transformation is characterized by a gradual change of the angle $\theta$ spanned by the magnetization and the $\mathbf{c}$-axis, taking place between $\approx40-130~K$. The zero-field MFM images across the SR range show strong variation in magnetic contrast.  Phenomenological analysis suggests that the SR and associated microstructural transformations are the results of competing anisotropy constants $K_1$ and $K_2$.

\section{Experimental details}

Single crystals of Fe$_3$Sn$_2$ have been grown by the chemical transport reaction method using preliminary synthesized polycrystalline material and iodine as a transport agent. The growth was carried out in a two-zone horizontal furnace in a temperature range between 670$^o$ and 740~$^o$C. Shiny hexagonal crystals with dimensions up to 4~mm along the largest side were obtained in the hot part of the ampule. The chemical composition of the samples was checked by energy-dispersive X-ray spectroscopy (EDS) analysis utilizing the ZEISS Crossbeam 550 system.

High-resolution x-ray diffraction (XRD) data were collected on the polycrystalline sample of Fe$_3$Sn$_2$. The measurements were performed in the transmission geometry at the ID22 beamline at the European Synchrotron Radiation Facility (ESRF, Grenoble, France) using the multianalyzer setup~\cite{Fitch2023, TSIRLIN2025}. The wavelength of 0.354217~\AA\ was calibrated with a silicon standard (NIST, 640c). The sample temperature was stabilized with the liquid helium-cooled cryostat. To reduce the effect of beam heating, the capillary was kept open at one end and exposed to the helium-flow atmosphere. The Rietveld refinement was performed using the JANA2006 program~\cite{jana2006}.

Magnetic properties were studied with the SQUID magnetometer (MPMS 3, Quantum Design) in the temperature range of $2-400$~K and magnetic fields up to 7~T. 

Neutron diffraction data were collected on the EIGER triple axis instrument (PSI, Switzerland) with a neutron wavelength of 2.36~\AA~\cite{Stuhr2017}. A few single crystals were coaligned to increase the total sample mass. The backscattering Laue x-ray showed that the relative misalignment between the individual crystals does not exceed 1~deg. The sample was inserted into the cryostat by its vertical [001] axis, placing the reciprocal $(HHL)$ plane in the horizontal scattering plane.
Graphite filter was placed after the sample to suppress the higher-order harmonics. 

To investigate the magnetic domain textures of Fe$_3$Sn$_2$, we perform MFM on an attocube attoAFM I, equipped with a superconducting magnet, and cooled by liquid Helium cryostat. The sample environment has ca. 150~mbar of He as an exchange gas. PPP-MFMR nanosensor probes were used; these have a hard magnetic coating and a radius of curvature of $\leq$ 50~nm. Once each temperature was reached, the equipment was allowed to thermalize for about an hour to ensure full thermal equilibrium. Data were collected in single pass with the tip lifted 200~nm above the defined contact point, and the pixel size is 20~nm. The images show changes in frequency, recorded via a phase-sensitive feedback loop, and, as usual, the image contrast is proportional to the gradient of the stray magnetic field at the tip ~\cite{Kazakova2019}.

\section{Results and discussion}

\begin{figure*} [t!]
 \centering
     \includegraphics[scale=0.5]{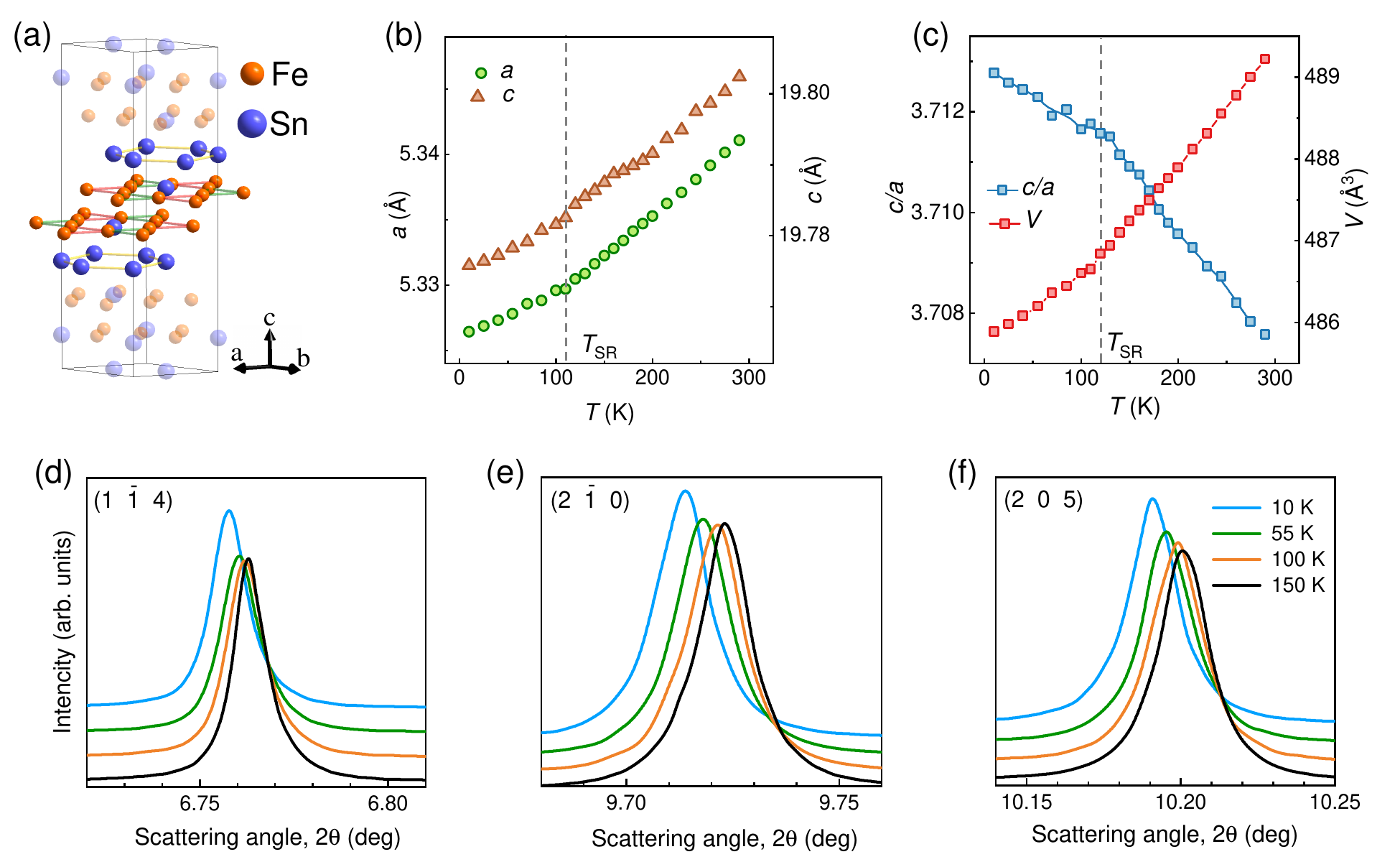}
     \caption{Crystal structure and high-resolution powder X-ray synchrotron diffraction for Fe$_3$Sn$_2$. (a) Schematic representation of the rhombohedral $R\overline{3}m$  crystal structure~\cite{Malaman1976}. (b) Temperature dependence of the lattice parameters $\mathbf{a}$ and $\mathbf{c}$. (c) Temperature variation of the ratio $\mathbf{c/a}$ and the unit-cell volume. The vertical dashed lines in (b) and (c) indicate the changes in the slope at the spin-reorientation temperature, T$_{SR}$. (d)-(f) Experimental profiles of the $(1\bar 14)$, $(2\bar 10)$, and $(205)$ Bragg reflections as measured at different temperatures by high-resolution synchrotron powder XRD. The same color code is used in panels (d-f).  }
     \label{fig:fig1XRD}
 \end{figure*}

Figures~\ref{fig:fig1XRD}(b) and (c) show the structural parameters derived from the Rietveld refinement of the XRD data between 300~K and 10~K. The cell parameter $\mathbf{a}$ decreases by ~$\sim3\%$, while $\mathbf{c}$ decreases by ~$\sim 2\%$ [see Fig.~\ref{fig:fig1XRD}(b)]. Both parameters exhibit a change in the slope below 120~K, which is also reflected in the temperature evolution of the $\mathbf{c/a}$ ratio and the unit-cell volume, Figure~\ref{fig:fig1XRD}(c), indicating a magnetoelastic response to the spin reorientation. Importantly, the reflection width changes only marginally between 300 and 10\,K, and none of the reflections splits on cooling [see Figs.~\ref{fig:fig1XRD}(d)-(e)]. Therefore, no deviation from the trigonal symmetry is observed down to the lowest measured temperatures.

\begin{figure*} [t!]
 \centering
     \includegraphics[scale=0.6]{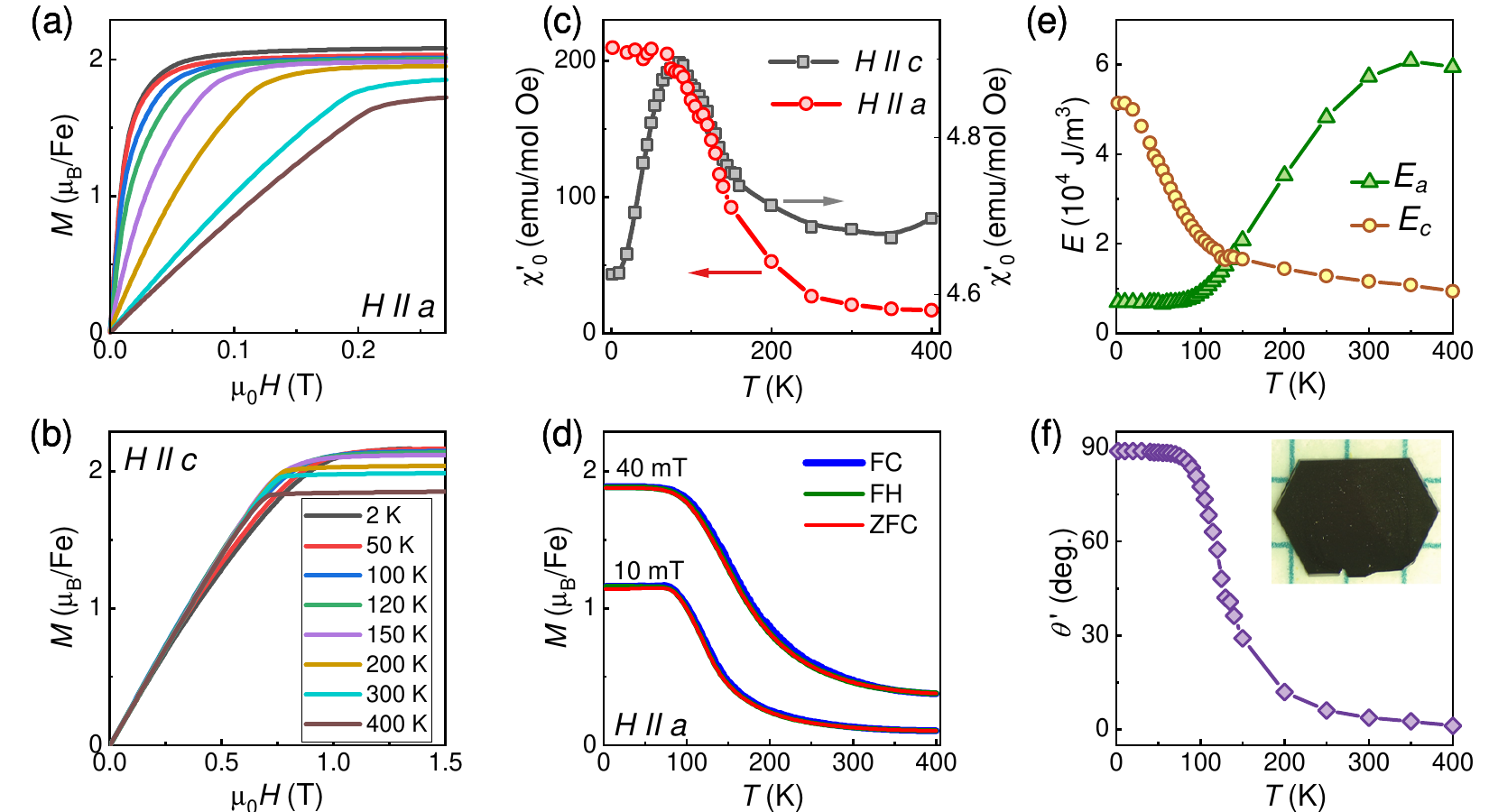}
     \caption{SQUID magnetometry data of ferromagnetic Fe$_3$Sn$_2$. (a) Magnetization curves \textit{vs} magnetic field applied along the $\mathbf{a}$ axis and (b) along the $\mathbf{c}$ axis, respectively. (c) Temperature dependence of differential magnetic susceptibility extracted from zero magnetic field measurements for $\mathbf{H||a}$ and $\mathbf{H||c}$. (d) Zero-field-cooled (ZFC), field-cooled (FC), and field-heated (FH) magnetization as a function of the temperature measured in fields of 10 and 40~mT applied along the $\mathbf{a}$ axis. (e) The temperature dependence of energy of magnetization for $E_a$ and $E_c$ used for calculating the uniaxial anisotropy constant $K_u$. (f) The temperature dependence of spin reorientation angle $\mathbf{\theta}'$ 
     calculated from the magnetization in $H_i=5~Oe$. The inset of (f) includes an optical image of the $\mathbf{ab}$ plane of the as-grown Fe$_3$Sn$_2$ crystal on a mm scale.}
     \label{fig:fig2SQUID}
 \end{figure*}

Figures~\ref{fig:fig2SQUID}(a) and (b) show the isothermal magnetization curves of Fe$_3$Sn$_2$ measured for magnetic fields applied in different orientations: within the $\mathbf{ab}$ plane, along the $\mathbf{a}$ axis (in-plane), and along the $\mathbf{c}$ axis (out-of-plane). At 2~K, full saturation of magnetization $M_s$ along the $\mathbf{a}$ axis is achieved in fields $\sim$0.1~T, while along $\mathbf{c}$ axis, the saturation requires approx. 1~T. The $\mathbf{M_s}$ calculated at 5~T are 2.09~$\mu_B$/Fe and 2.17~$\mu_B$/Fe for $\mathbf{H||a}$ and $\mathbf{H||c}$, respectively. With increasing temperature, the saturation field increases for $\mathbf{H||a}$ and decreases for $\mathbf{H||c}$. The initial susceptibility extracted from fields close to zero, ${\chi'_0}$, measured for $\mathbf{H||a}$ and $\mathbf{H||c}$ is shown in Figure~\ref{fig:fig2SQUID}(c).

The temperature-dependent magnetization for $\mathbf{H||a}$ was measured in various fields and cooling protocols. Figure~\ref{fig:fig2SQUID}(d)  shows a continuous increase of magnetization as the temperature decreases to 60~K, saturating below this temperature. Unlike the findings in Ref. ~\cite{Heritage2020}, our data show gradual variation with temperature without abrupt changes between 400~K and 2~K. The zero field-cooled magnetization (ZFC) and field-cooled magnetization (FC) measurements display only minor hysteresis, which can be likely attributed to the dynamics of magnetic microstructures described later.

Next, we discuss the uniaxial crystal anisotropy energy density (E$_A$) that for hexagonal system is usually expressed as a power series of the form~\cite{Ohandley1999modern}:
\begin{align}\label{eq:E_A}
   E_A = \sum_{n} K_{n} \sin^{2n}\theta \ .
\end{align}
Here, $K_n$ is the uniaxial anisotropy constant of order $n$ and $\mathbf{\theta}$ is the angle between $M$ and the $\mathbf{c}$ axis. 
In most cases, it is sufficient to consider only the first two terms:
\begin{align}\label{eq:E_A2}
 E_A=K_0+K_1 \sin^2 \theta+K_2\sin^4\theta+...
\end{align}

\begin{figure} [h]
 \centering
     \includegraphics[scale=0.45]{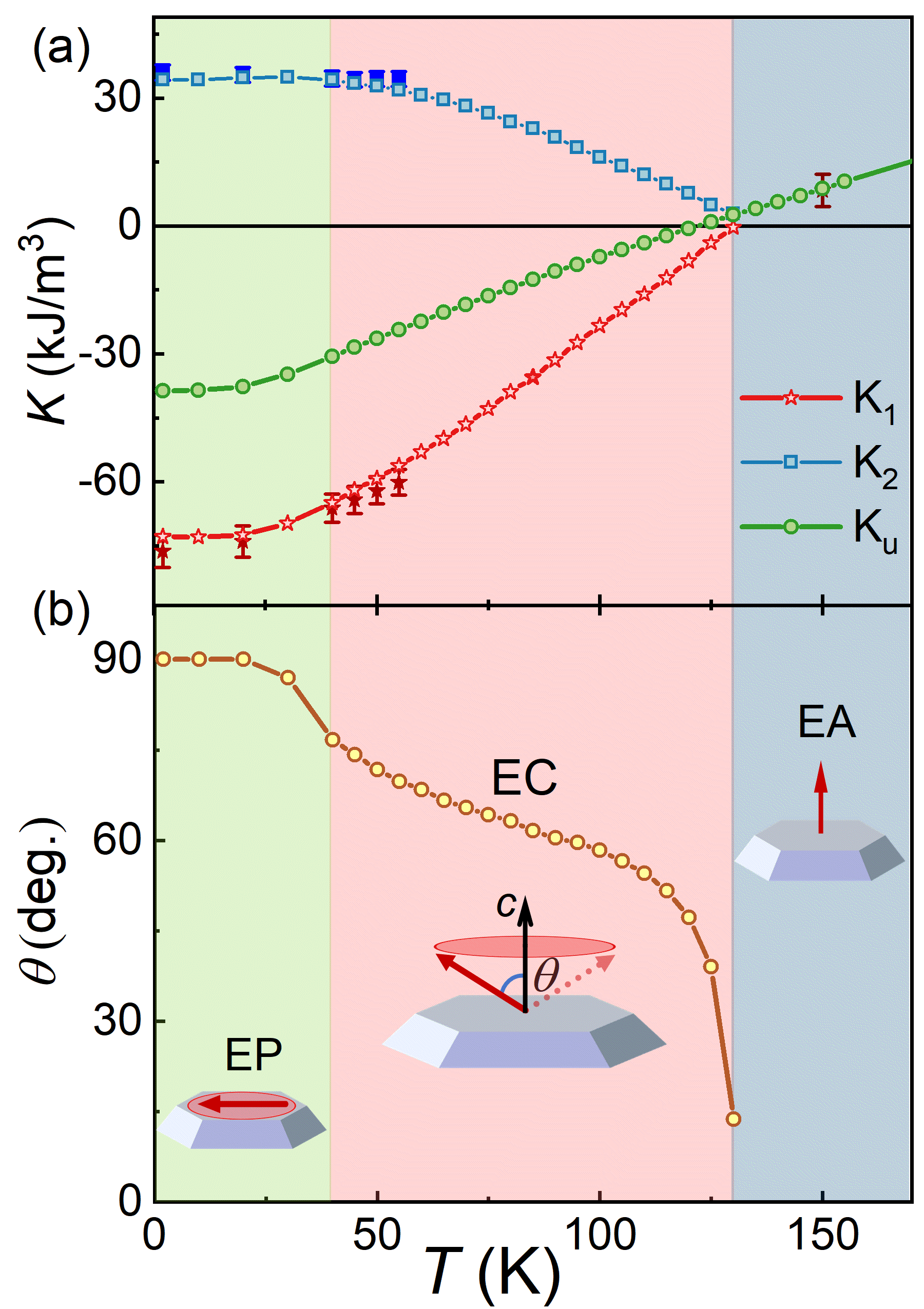}
     \caption{(a) Temperature dependence of the anisotropy constants $K_u$, $K_1$ and $K_2$ obtained by the area method and eq.\ref{eq:Suscept} (open symbols). The closed symbols show $K_1$ and $K_2$ obtained by the Sucksmith-Thompson method. (b) Temperature dependence of the spin reorientation angle  $\mathbf{\theta}$, as calculated using $K_1$ and $K_2$. The insets show schematically the spin arrangement in easy-plane (EP), easy-cone (EC), and easy-axis states (EA).} 
     \label{fig:fig3K}
 \end{figure}

The anisotropy energy minimization in the presence of an external applied field provides the equilibrium direction of magnetization. In the limit of zero external field, the total anisotropy energy is $K_u\approx K_1+K_2$. This $K_u$ constant can be estimated directly from the experimental data as the difference between the energy necessary to saturate the magnetization along the two orthogonal directions:

Figure~\ref{fig:fig2SQUID}(e) presents the temperature dependence of ${E_a}$ and ${E_c}$ used to calculate the uniaxial anisotropy constant $K_u$. Here, the magnetization was corrected for the demagnetization coefficient,  $H_i=H-DM$, where $H$ is the applied external magnetic field, $D$ represents the demagnetization coefficient, and $M$ is the magnetization. $E_a$ decreases with a decrease in temperature, while $E_c$ increases, reflecting the change of the anisotropy. At 120~K, the integrals have equal values, indicating that the same energy is required to align the magnetic moment along the $\mathbf{a}$ or $\mathbf{c}$ axes. Consequently, $K_u=0$ at this temperature [see Fig.~\ref{fig:fig3K}(a)]. At temperatures below 120~K, ${K_u}$ is negative, reaching $-45$~kJ/m$^3$ at 2~K, while above is positive and reaches $57$~kJ/m$^3$ at 400~K.

 Figure~\ref{fig:fig3K}(a) shows the temperature dependence of the anisotropy constants $K_u$, $K_1$, and $K_2$. In the temperature range where the orientation of moments is well-defined, we also calculated $K_1$ and $K_2$ independently by the Sucksmith-Thompson (S-T) method fitting the S-T isotherms $M/H=f(M^2)$~\cite{Sucksmith1954}. At 2~K, $K_1$ is negative (easy-plane anisotropy) and reaches $\sim-75$~kJ/m$^3$ [see Figure~\ref{fig:fig3K}(a)]. At high temperatures, $K_1$ is positive (easy-axis anisotropy) and reaches $\sim55$~kJ/m$^3$ at 400~K. In contrast, $K_2$ remains positive throughout the whole temperature range. At 2~K, $K_2$ has a value $\sim30$~kJ/m$^3$. At the intermediate temperatures, 50~K $\leq$~T~$\leq$ 150~K, the S-T isotherms show strong nonlinearity and this method gives non-reliable values anisotropy constants.

In the following, we present a new approach for extracting anisotropy constants when the S-T isotherms become nonlinear. The minimization of magnetic free energy described in Eq.~\ref{eq:E_A}, with respect to the reorientation angle $\mathbf{\theta}$ (being the order parameter), gives the equilibrium configuration. In this case, the relation between the zero-field differential susceptibility for $\mathbf{H||c}$ and the anisotropy constants, the saturation magnetization for $\mathbf{H||c}$, and the demagnetization has the following form (see Appendix): 
 
 \begin{align}\label{eq:Suscept}
\chi{(T)}=\dfrac{M_s^2(T)}{D~M_s^2(T)-2(K_1+2K_2)}    
\end{align}
The resulting temperature dependence of $K_1$ and $K_2$ is shown in Figure~\ref{fig:fig3K}(a). Below 50~K, the values of $K_1$ and $K_2$ agree well with those obtained by the S-T method, corroborating the data obtained through the new approach.

The interplay between $K_1$ and $K_2$ controls the transition from the easy axis (EA) to the easy plane (EP) via the easy-cone (EC) state~\cite{Cullity2008}. In the EC state, the easy direction of the magnetization is tilted away both from the hexagonal axis and the hexagonal plane, i.e. $0<\theta<90^{\circ}$. When the anisotropy in the hexagonal plane is negligible, the easy axes lie on the surface of a cone. In the EC state, the rotation angle is defined by $\theta(T)={\rm arcsin}\sqrt{-K_1/2K_2}$ ~\cite{Algarabel1988}. 

The temperature dependence of $\mathbf{\theta}$ calculated from the anisotropy constants $K_1$ and $K_2$ is shown in Figure~\ref{fig:fig3K}(b). The transition from EP to EC takes place at $T_{SR1}\approx40$~ K, when condition $-K_1=2K_2$ is satisfied. Upon heating in the easy-cone state, the magnetic moments rotate gradually from the $\mathbf{a}$ axis towards the $\mathbf{c}$ axis. At $T_{SR2}\approx130$~K, $K_1$ changes sign, defining the transition to the easy-axis state.

The angle of SR was additionally calculated from the ratio of low-field magnetization for $\mathbf{H||a}$ and $\mathbf{H||c}$ for internal magnetic field $H_i$=5~Oe using the relation $\theta'={\rm arctan}({M_a/M_c})$ [see Figure~\ref{fig:fig2SQUID}(f)]. The calculated  $\theta'$ also shows gradual decrease with increasing temperature, being in qualitative agreement with the results using anisotropy constants.

For an independent quantification of the SR angle, we performed single-crystal neutron diffraction experiments. According to the neutron polarization factor, the scattering intensity vanishes when the neutron momentum transfer $\mathbf{Q}$ aligns with the magnetic structure factor $\mathbf{M_Q}$. Therefore, it is convenient to track the in-plane projection of the spin upon its canting by the intensity of the allowed Bragg peaks of the indexes $(00L)$, where $L = 3, 6, 9\ldots, 3n$. Indeed, when the spins are strictly oriented along the $\mathbf{c}$ axis, the $(00L)$ peaks contain only the nuclear intensity and no magnetic contribution. As soon as the spins deviate from the $\mathbf{c}$ axis, the $\mathbf{Q} \parallel \mathbf{M_Q}$ condition is no longer fulfilled, and a finite magnetic intensity $\propto S_{ab}$ appears on top of the nuclear Bragg peaks.

\begin{figure} [h!]
 \centering
     \includegraphics[width=0.8\linewidth]{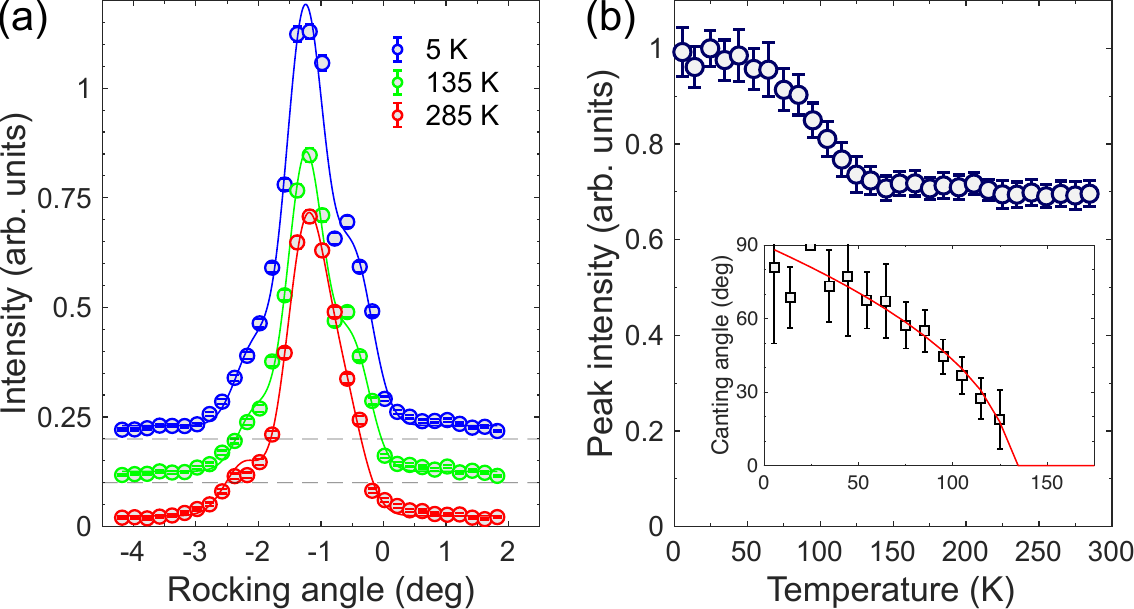}
     \caption{Neutron diffraction: (a) The rocking-scan intensity of the (003) Bragg peak at different temperatures. The data were offset for clarity by the dashed lines. (b) The temperature dependence of the integrated intensity (area) of the (003) Bragg peak. The inset shows the calculated spin reorientation angle $\theta$. The red curve is a guide to an eye. }
     \label{fig:fig4Neutron}
 \end{figure}

Figure~\ref{fig:fig4Neutron}(a) shows the intensity of the Bragg peak [003] collected by rotating the crystal around its [001] at a fixed detector angle $2\theta=41.21^o$ (rocking scan). As can be seen, the measurements reveal almost identical Bragg peaks at 285~K and 135~K,  whereas the peak intensity significantly increases at a lower temperature of 5~K. To analyze its temperature evolution, we plotted the total intensity integrated over the rocking angle of the [003] peak in Fig.~\ref{fig:fig4Neutron}(b). The flat curve from room temperature down to $\sim$130~K suggests the spins remain along the $c$ axis and only the temperature-independent nuclear intensity is recorded. Below this temperature, the intensity shows a gradual increase over a wide temperature range down to $\sim$40~K, after which it saturates. Assuming that the spins are fully in-plane at the lowest temperature, we can follow the temperature dependence of the SR by the $\mathbf{\theta}$ angle ($\mathbf{\theta}=0$ when $\mathbf{M} \parallel c$) estimated from neutron diffraction, as shown in the inset of Fig.~\ref{fig:fig4Neutron}(b). This rotation of magnetic moments from the $\mathbf{c}$ axis to the $\mathbf{a}$ axis should lower the symmetry by breaking the three-fold rotational symmetry for any $\mathbf{\theta}>0$. The fact that the symmetry lowering from the trigonal state to an orthorhombic state has not been observed by our high-resolution x-ray study may be due to a) the weakness of the corresponding distortion and/or b) the restoring of the trigonal symmetry in macroscopic sense due to the coexistence of the energetically equivalent six domains of the EP state.

\begin{figure} [h]
 \centering
     \includegraphics[scale=1]{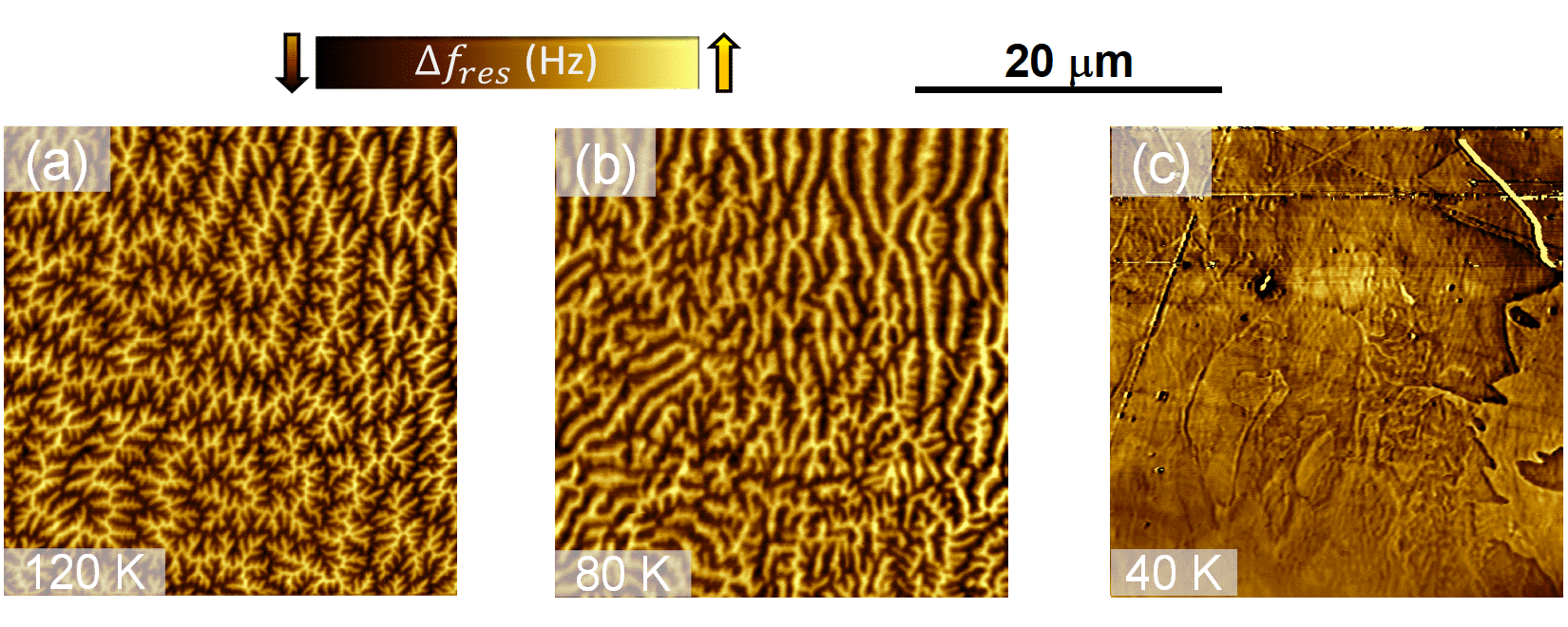}
     \caption{Evolution of the magnetic domain patterns illustrated using MFM contrasts.
     Image (a) shows dendrite-like domains with out-of-plane component of magnetization measured at 120~K. 
     Panel (b) shows a stripe domain pattern forming at 80~K. Panel (c) illustrates the lack of out-of-plane contrast at 40~K.}
     \label{fig:fig5MFM}
 \end{figure}

For an independent confirmation of the SR in our Fe$_3$Sn$_2$ single crystals, the magnetic domain pattern were imaged by MFM in zero field on the as-grown $\mathbf{ab}$  plane of the crystal shown in the inset of Figure~\ref{fig:fig2SQUID}(f). Figure~\ref{fig:fig5MFM}(a) shows the MFM pattern at 120~K, revealing a dendrite-like fern contrast. This is similar to the domain pattern at 300~K which previous work showed arises from magnetization pointing upward and downward along the $\mathbf{c}$  axis, see e.g. Refs. ~\cite{Fayyazi2019, Heritage2020, Altthaler2021}. We note that the fern pattern is caused by surface branching of the underlying domains, and therefore the domain patterns observed here with MFM in Fe$_3$Sn$_2$ do not reflect the magnetic microstructure in the bulk. Indeed, such dendrite-like domain patterns have been reported to be characteristic of uniaxial easy-axis magnets ~\cite{Hubert2009}. With cooling down to 80~K, where the magnetic easy axis is already tilted away from the $c$ axis according to our magnetometry and neutron diffraction results, the MFM contrast undergoes gradual changes, namely the dendrite-like structure transforms to a more simple stripe domain pattern. Similar transformation of magnetic domains in Fe$_3$Sn$_2$ thin plates was reported to result from tilting of spins away from the $\mathbf{c}$ axis~\cite{Chen2021}. With further cooling to 40~K the stripe pattern vanishes. In fact, in Figure~\ref{fig:fig5MFM}(c) the magnetic contrast caused by out-of-plane stray fields disappears, indicating that the sample becomes in-plane magnetized. In that case, vortices and antivotices (also termed in-plane flux closure domains) are expected to form, as shown for the easy-plane ferromagnet Fe$_3$Sn~\cite{Prodan2023}. These results are qualitatively in agreement with MFM results reported in Refs.~\cite{Heritage2020, Xie2024} and clearly demonstrate the onset of a SR from a high-temperature easy-axis state to a low-temperature easy-plane state.

A quantitative analysis of the magnetic behavior at the micrometer-length scale is usually obtained by micromagnetic simulations, which can provide a comprehensive description of MFM images. In combination with density functional theory, these simulations can consider various forms of exchange interactions and anisotropies. 
While here we focus on the analysis of the temperature-dependent SR process, MFM imaging is only used to provide another independent confirmation of the SR. A multi-scale approach reproducing the magnetic patterns observed in zero as well as finite magnetic fields will be presented in a forthcoming work.
 
\section{CONCLUSION}

In conclusion, we have investigated the temperature-driven spin reorientation in the topological kagome ferromagnet Fe$_3$Sn$_2$ by high-resolution synchrotron x-ray diffraction, magnetometry, neutron diffraction, and MFM techniques supported by phenomenological analysis of magnetic anisotropy. Our study reveals an intermediate easy-cone state that appears during the reorientation from the easy-plane towards the easy-axis state. Although spin reorientation is accompanied by a weak magnetoelastic effect, no lowering of the crystal symmetry from trigonal was resolved down to the lowest temperatures. The temperature dependence of the first- and second-order anisotropy constants, $K_1$ and $K_2$, and the evolution of the magnetic structure factor $\mathbf{M_Q}(T)$ show that the easy-cone state is stable between $T_{SR1}\approx40$~K and $T_{SR2}\approx130$~K. The spin reorientation angle $\theta$, determined from the magnetometry and neutron diffraction data, exhibits a gradual change between $T_{SR1}$ and $T_{SR2}$. Furthermore, zero-field MFM images across the SR range evidence a transformation in magnetic contrast, from a dendritic structure at 120~K to well-defined domain walls at 40~K, providing a direct visualization of spin reorientation in Fe$_3$Sn$_2$. Our phenomenological analysis highlights the dominant role of competition between anisotropy constants.
These findings provide a clear insight into the nature and dynamics of spin reorientation in topological magnet Fe$_3$Sn$_2$, highlighting its fundamental role in the evolution of magnetic anisotropy.

\begin{acknowledgments}
We thank ESRF for providing the beamtime for the project HC-4369 and acknowledge Ola Grendal as well as Andrew Fitch for their technical support at ID22. This work was supported by the Deutsche Forschungsgemeinschaft (DFG, German Research Foundation) -- TRR 360 -- 492547816. D.M.E. acknowledges and thank the DFG for financial support via DFG individual fellowship number EV 305/1-1. VT acknowledges support through project ANCD (code 011201, Moldova).
\end{acknowledgments}

\include{append}

\bibliography{Main}

\end{document}

%% file: append.tex
\appendix
\section{The formula of differential susceptibility}
In the following we present a brief description of the derivation of Eq.~\eqref{eq:Suscept}.
To simplify formulas, we introduce the reduced parameters: $k=K_1/K_2$, $A=M_sH/K_2$, $B=DM_s^2/(2K_2)$.
The free energy is a functional of applied field and temperature and contains besides the Zeeman term the demagnetization energy contribution: 

\begin{widetext}
\begin{align}
    E[k, A, B] = \frac{E-K_0}{K_2} = k \sin^2 \theta  + \sin^4 \theta  -A \cos \theta  + B \cos^2 \theta  
\end{align}
\end{widetext}
The substitution with a variable $\gamma=cos \ \theta = M_{||}/M_s $, leads the free energy into a fourth-order expression in $\gamma$. The equilibrium configuration for the orientation of the magnetic moments is found by minimization with respect to the reorientation angle $\theta = Arccos \ \gamma$. Conventionally, this angle is measured with respect to the $c$ axis. The partial derivative with respect to $\gamma$: $\partial E[k, A,B]/\partial \gamma =0$, therefore, produces a third-order equation that is analytically solvable. Two of its complex solutions are discarded, and the real solution has an analytical form:
\begin{widetext}
\begin{align} \nonumber
    \gamma= \frac{\left(432 A+\sqrt{186624 A^2+4 (-48+24 B-24 k)^3}\right)^{1/3}}{12\cdot 2^{1/3}}-\\
-{\frac{-48+24 B-24 k}{6\cdot  2^{2/3} \left(432 A+\sqrt{186624 A^2+4 (-48+24 B-24 k)^3}\right)^{1/3}}}\
\end{align}
\end{widetext}

The above equation represents in fact the reduced magnetization $M_{||}/M_s$, with $M_{||}$ being the magnetization along the applied field and $M_s$ the saturated magnetization (as described in the main text) . Accordingly, the differential susceptibility is computed as a derivative of magnetization with respect to the applied field:
$$\chi(H,T) = \frac{dM_{||}}{dH} = \frac{M^2_s}{K_2} \frac{d\gamma}{dA} $$
Within the reduced variables taking the derivative with respect to the variable $A$: $d\gamma/dA$, finally the limit $A\rightarrow 0$ gives:
\begin{align}
    \chi(k, A, B) \vert_{A \rightarrow 0}= \frac{M^2_s}{K_2} \frac{1}{2(-2+B-k)} 
\end{align}
Inserting back the forms of the reduced variables $B$ and $k$ leads to the form presented as Eq.~\eqref{eq:Suscept}.

%% file: Main.bbl
\begin{thebibliography}{75}%
\makeatletter
\providecommand \@ifxundefined [1]{%
 \@ifx{#1\undefined}
}%
\providecommand \@ifnum [1]{%
 \ifnum #1\expandafter \@firstoftwo
 \else \expandafter \@secondoftwo
 \fi
}%
\providecommand \@ifx [1]{%
 \ifx #1\expandafter \@firstoftwo
 \else \expandafter \@secondoftwo
 \fi
}%
\providecommand \natexlab [1]{#1}%
\providecommand \enquote  [1]{``#1''}%
\providecommand \bibnamefont  [1]{#1}%
\providecommand \bibfnamefont [1]{#1}%
\providecommand \citenamefont [1]{#1}%
\providecommand \href@noop [0]{\@secondoftwo}%
\providecommand \href [0]{\begingroup \@sanitize@url \@href}%
\providecommand \@href[1]{\@@startlink{#1}\@@href}%
\providecommand \@@href[1]{\endgroup#1\@@endlink}%
\providecommand \@sanitize@url [0]{\catcode `\\12\catcode `\$12\catcode
  `\&12\catcode `\#12\catcode `\^12\catcode `\_12\catcode `\%12\relax}%
\providecommand \@@startlink[1]{}%
\providecommand \@@endlink[0]{}%
\providecommand \url  [0]{\begingroup\@sanitize@url \@url }%
\providecommand \@url [1]{\endgroup\@href {#1}{\urlprefix }}%
\providecommand \urlprefix  [0]{URL }%
\providecommand \Eprint [0]{\href }%
\providecommand \doibase [0]{http://dx.doi.org/}%
\providecommand \selectlanguage [0]{\@gobble}%
\providecommand \bibinfo  [0]{\@secondoftwo}%
\providecommand \bibfield  [0]{\@secondoftwo}%
\providecommand \translation [1]{[#1]}%
\providecommand \BibitemOpen [0]{}%
\providecommand \bibitemStop [0]{}%
\providecommand \bibitemNoStop [0]{.\EOS\space}%
\providecommand \EOS [0]{\spacefactor3000\relax}%
\providecommand \BibitemShut  [1]{\csname bibitem#1\endcsname}%
\let\auto@bib@innerbib\@empty
\bibitem [{\citenamefont {Buschow}\ and\ \citenamefont
  {de~Boer}(2004)}]{Buschow2003}%
  \BibitemOpen
  \bibfield  {author} {\bibinfo {author} {\bibfnamefont {K.H.J.}\ \bibnamefont
  {Buschow}}\ and\ \bibinfo {author} {\bibfnamefont {F.R.}\ \bibnamefont
  {de~Boer}},\ }\href@noop {} {\emph {\bibinfo {title} {{Physics of Magnetism
  and Magnetic Materials}}}},\ \bibinfo {edition} {1st}\ ed.\ (\bibinfo
  {publisher} {Kluwer Academic Publishers},\ \bibinfo {address} {New York,
  NY},\ \bibinfo {year} {2004})\BibitemShut {NoStop}%
\bibitem [{\citenamefont {Dzyaloshinsky}(1958)}]{Dzyaloshinsky1958}%
  \BibitemOpen
  \bibfield  {author} {\bibinfo {author} {\bibfnamefont {I.}~\bibnamefont
  {Dzyaloshinsky}},\ }\bibfield  {title} {\enquote {\bibinfo {title} {{A
  thermodynamic theory of “weak” ferromagnetism of antiferromagnetics}},}\
  }\href {\doibase 10.1016/0022-3697(58)90076-3} {\bibfield  {journal}
  {\bibinfo  {journal} {Journal of Physics and Chemistry of Solids}\ }\textbf
  {\bibinfo {volume} {4}},\ \bibinfo {pages} {241--255} (\bibinfo {year}
  {1958})}\BibitemShut {NoStop}%
\bibitem [{\citenamefont {Moriya}(1960)}]{Moriya1960}%
  \BibitemOpen
  \bibfield  {author} {\bibinfo {author} {\bibfnamefont {Tôru}\ \bibnamefont
  {Moriya}},\ }\bibfield  {title} {\enquote {\bibinfo {title} {{Anisotropic
  Superexchange Interaction and Weak Ferromagnetism}},}\ }\href {\doibase
  10.1103/physrev.120.91} {\bibfield  {journal} {\bibinfo  {journal} {Physical
  Review}\ }\textbf {\bibinfo {volume} {120}},\ \bibinfo {pages} {91--98}
  (\bibinfo {year} {1960})}\BibitemShut {NoStop}%
\bibitem [{\citenamefont {Locatelli}\ \emph {et~al.}(2013)\citenamefont
  {Locatelli}, \citenamefont {Cros},\ and\ \citenamefont
  {Grollier}}]{Locatelli2013}%
  \BibitemOpen
  \bibfield  {author} {\bibinfo {author} {\bibfnamefont {N.}~\bibnamefont
  {Locatelli}}, \bibinfo {author} {\bibfnamefont {V.}~\bibnamefont {Cros}}, \
  and\ \bibinfo {author} {\bibfnamefont {J.}~\bibnamefont {Grollier}},\
  }\bibfield  {title} {\enquote {\bibinfo {title} {{Spin-torque building
  blocks}},}\ }\href {\doibase 10.1038/nmat3823} {\bibfield  {journal}
  {\bibinfo  {journal} {Nature Materials}\ }\textbf {\bibinfo {volume} {13}},\
  \bibinfo {pages} {11--20} (\bibinfo {year} {2013})}\BibitemShut {NoStop}%
\bibitem [{\citenamefont {Zuo}\ \emph {et~al.}(2021)\citenamefont {Zuo},
  \citenamefont {Liu}, \citenamefont {Qiao}, \citenamefont {Zhang},
  \citenamefont {Chen}, \citenamefont {Su}, \citenamefont {Liu}, \citenamefont
  {Cao}, \citenamefont {Zhao}, \citenamefont {Wang}, \citenamefont {Hu},
  \citenamefont {Sun}, \citenamefont {Jiang},\ and\ \citenamefont
  {Shen}}]{Zuo2021}%
  \BibitemOpen
  \bibfield  {author} {\bibinfo {author} {\bibfnamefont {Shulan}\ \bibnamefont
  {Zuo}}, \bibinfo {author} {\bibfnamefont {Jun}\ \bibnamefont {Liu}}, \bibinfo
  {author} {\bibfnamefont {Kaiming}\ \bibnamefont {Qiao}}, \bibinfo {author}
  {\bibfnamefont {Ying}\ \bibnamefont {Zhang}}, \bibinfo {author}
  {\bibfnamefont {Jie}\ \bibnamefont {Chen}}, \bibinfo {author} {\bibfnamefont
  {Na}~\bibnamefont {Su}}, \bibinfo {author} {\bibfnamefont {Yanli}\
  \bibnamefont {Liu}}, \bibinfo {author} {\bibfnamefont {Jun}\ \bibnamefont
  {Cao}}, \bibinfo {author} {\bibfnamefont {Tongyun}\ \bibnamefont {Zhao}},
  \bibinfo {author} {\bibfnamefont {Jingmin}\ \bibnamefont {Wang}}, \bibinfo
  {author} {\bibfnamefont {Fengxia}\ \bibnamefont {Hu}}, \bibinfo {author}
  {\bibfnamefont {Jirong}\ \bibnamefont {Sun}}, \bibinfo {author}
  {\bibfnamefont {Chengbao}\ \bibnamefont {Jiang}}, \ and\ \bibinfo {author}
  {\bibfnamefont {Baogen}\ \bibnamefont {Shen}},\ }\bibfield  {title} {\enquote
  {\bibinfo {title} {{Spontaneous Topological Magnetic Transitions in NdCo$_5$
  Rare‐Earth Magnets}},}\ }\href {\doibase 10.1002/adma.202103751} {\bibfield
   {journal} {\bibinfo  {journal} {Advanced Materials}\ }\textbf {\bibinfo
  {volume} {33}} (\bibinfo {year} {2021}),\ 10.1002/adma.202103751}\BibitemShut
  {NoStop}%
\bibitem [{\citenamefont {Kisi\ifmmode~\check{c}\else \v{c}\fi{}ek}\ \emph
  {et~al.}(2024)\citenamefont {Kisi\ifmmode~\check{c}\else \v{c}\fi{}ek},
  \citenamefont {Dominko}, \citenamefont {\ifmmode~\check{C}\else
  \v{C}\fi{}ulo}, \citenamefont {Rapljenovi\ifmmode~\acute{c}\else \'{c}\fi{}},
  \citenamefont {Kuve\ifmmode \check{z}\else
  \v{z}\fi{}di\ifmmode~\acute{c}\else \'{c}\fi{}}, \citenamefont {Dragi\ifmmode
  \check{c}\else \v{c}\fi{}evi\ifmmode~\acute{c}\else \'{c}\fi{}},
  \citenamefont {Berger}, \citenamefont {Rocquefelte}, \citenamefont {Herak},\
  and\ \citenamefont {Ivek}}]{Kisi2024}%
  \BibitemOpen
  \bibfield  {author} {\bibinfo {author} {\bibfnamefont {Virna}\ \bibnamefont
  {Kisi\ifmmode~\check{c}\else \v{c}\fi{}ek}}, \bibinfo {author} {\bibfnamefont
  {Damir}\ \bibnamefont {Dominko}}, \bibinfo {author} {\bibfnamefont {Matija}\
  \bibnamefont {\ifmmode~\check{C}\else \v{C}\fi{}ulo}}, \bibinfo {author}
  {\bibfnamefont {\ifmmode \check{Z}\else~\v{Z}\fi{}eljko}\ \bibnamefont
  {Rapljenovi\ifmmode~\acute{c}\else \'{c}\fi{}}}, \bibinfo {author}
  {\bibfnamefont {Marko}\ \bibnamefont {Kuve\ifmmode \check{z}\else
  \v{z}\fi{}di\ifmmode~\acute{c}\else \'{c}\fi{}}}, \bibinfo {author}
  {\bibfnamefont {Martina}\ \bibnamefont {Dragi\ifmmode \check{c}\else
  \v{c}\fi{}evi\ifmmode~\acute{c}\else \'{c}\fi{}}}, \bibinfo {author}
  {\bibfnamefont {Helmuth}\ \bibnamefont {Berger}}, \bibinfo {author}
  {\bibfnamefont {Xavier}\ \bibnamefont {Rocquefelte}}, \bibinfo {author}
  {\bibfnamefont {Mirta}\ \bibnamefont {Herak}}, \ and\ \bibinfo {author}
  {\bibfnamefont {Tomislav}\ \bibnamefont {Ivek}},\ }\bibfield  {title}
  {\enquote {\bibinfo {title} {{Spin-Reorientation-Driven Linear
  Magnetoelectric Effect in Topological Antiferromagnet
  ${\mathrm{Cu}}_{3}{\mathrm{TeO}}_{6}$}},}\ }\href {\doibase
  10.1103/PhysRevLett.132.096701} {\bibfield  {journal} {\bibinfo  {journal}
  {Phys. Rev. Lett.}\ }\textbf {\bibinfo {volume} {132}},\ \bibinfo {pages}
  {096701} (\bibinfo {year} {2024})}\BibitemShut {NoStop}%
\bibitem [{\citenamefont {Pal}\ \emph {et~al.}(2024)\citenamefont {Pal},
  \citenamefont {Pal}, \citenamefont {Mondal}, \citenamefont {Sharma},
  \citenamefont {Das}, \citenamefont {Mandal},\ and\ \citenamefont
  {Pal}}]{Pal2024}%
  \BibitemOpen
  \bibfield  {author} {\bibinfo {author} {\bibfnamefont {Riju}\ \bibnamefont
  {Pal}}, \bibinfo {author} {\bibfnamefont {Buddhadeb}\ \bibnamefont {Pal}},
  \bibinfo {author} {\bibfnamefont {Suchanda}\ \bibnamefont {Mondal}}, \bibinfo
  {author} {\bibfnamefont {Rajesh~O.}\ \bibnamefont {Sharma}}, \bibinfo
  {author} {\bibfnamefont {Tanmoy}\ \bibnamefont {Das}}, \bibinfo {author}
  {\bibfnamefont {Prabhat}\ \bibnamefont {Mandal}}, \ and\ \bibinfo {author}
  {\bibfnamefont {Atindra~Nath}\ \bibnamefont {Pal}},\ }\bibfield  {title}
  {\enquote {\bibinfo {title} {{Spin-reorientation driven emergent phases and
  unconventional magnetotransport in quasi-2D vdW ferromagnet
  Fe$_4$GeTe$_2$}},}\ }\href {\doibase 10.1038/s41699-024-00463-y} {\bibfield
  {journal} {\bibinfo  {journal} {npj 2D Materials and Applications}\ }\textbf
  {\bibinfo {volume} {8}} (\bibinfo {year} {2024}),\
  10.1038/s41699-024-00463-y}\BibitemShut {NoStop}%
\bibitem [{\citenamefont {Johnson}\ \emph {et~al.}(1996)\citenamefont
  {Johnson}, \citenamefont {Bloemen}, \citenamefont {den Broeder},\ and\
  \citenamefont {de~Vries}}]{Johnson1996}%
  \BibitemOpen
  \bibfield  {author} {\bibinfo {author} {\bibfnamefont {M.~T.}\ \bibnamefont
  {Johnson}}, \bibinfo {author} {\bibfnamefont {P.~J.~H.}\ \bibnamefont
  {Bloemen}}, \bibinfo {author} {\bibfnamefont {F.~J.~A.}\ \bibnamefont {den
  Broeder}}, \ and\ \bibinfo {author} {\bibfnamefont {J.~J.}\ \bibnamefont
  {de~Vries}},\ }\bibfield  {title} {\enquote {\bibinfo {title} {Magnetic
  anisotropy in metallic multilayers},}\ }\href {\doibase
  10.1088/0034-4885/59/11/002} {\bibfield  {journal} {\bibinfo  {journal}
  {Reports on Progress in Physics}\ }\textbf {\bibinfo {volume} {59}},\
  \bibinfo {pages} {1409--1458} (\bibinfo {year} {1996})}\BibitemShut {NoStop}%
\bibitem [{\citenamefont {White}(1969)}]{White1969}%
  \BibitemOpen
  \bibfield  {author} {\bibinfo {author} {\bibfnamefont {R.~L.}\ \bibnamefont
  {White}},\ }\bibfield  {title} {\enquote {\bibinfo {title} {Review of recent
  work on the magnetic and spectroscopic properties of the rare-earth
  orthoferrites},}\ }\href {\doibase 10.1063/1.1657530} {\bibfield  {journal}
  {\bibinfo  {journal} {Journal of Applied Physics}\ }\textbf {\bibinfo
  {volume} {40}},\ \bibinfo {pages} {1061--1069} (\bibinfo {year}
  {1969})}\BibitemShut {NoStop}%
\bibitem [{\citenamefont {Yamaguchi}(1974)}]{Yamaguchi1974}%
  \BibitemOpen
  \bibfield  {author} {\bibinfo {author} {\bibfnamefont {T.}~\bibnamefont
  {Yamaguchi}},\ }\bibfield  {title} {\enquote {\bibinfo {title} {{Theory of
  spin reorientation in rare-earth orthochromites and orthoferrites}},}\ }\href
  {\doibase 10.1016/s0022-3697(74)80003-x} {\bibfield  {journal} {\bibinfo
  {journal} {Journal of Physics and Chemistry of Solids}\ }\textbf {\bibinfo
  {volume} {35}},\ \bibinfo {pages} {479--500} (\bibinfo {year}
  {1974})}\BibitemShut {NoStop}%
\bibitem [{\citenamefont {Algarabel}\ \emph {et~al.}(1988)\citenamefont
  {Algarabel}, \citenamefont {Moral}, \citenamefont {Ibarra},\ and\
  \citenamefont {Arnaudas}}]{Algarabel1988}%
  \BibitemOpen
  \bibfield  {author} {\bibinfo {author} {\bibfnamefont {P.A.}\ \bibnamefont
  {Algarabel}}, \bibinfo {author} {\bibfnamefont {A.del}\ \bibnamefont
  {Moral}}, \bibinfo {author} {\bibfnamefont {M.R.}\ \bibnamefont {Ibarra}}, \
  and\ \bibinfo {author} {\bibfnamefont {J.I.}\ \bibnamefont {Arnaudas}},\
  }\bibfield  {title} {\enquote {\bibinfo {title} {{Spin reorientation in
  RECo$_5$ compounds: A.C. susceptibility and thermal expansion}},}\ }\href
  {\doibase 10.1016/0022-3697(88)90053-4} {\bibfield  {journal} {\bibinfo
  {journal} {Journal of Physics and Chemistry of Solids}\ }\textbf {\bibinfo
  {volume} {49}},\ \bibinfo {pages} {213--222} (\bibinfo {year}
  {1988})}\BibitemShut {NoStop}%
\bibitem [{\citenamefont {Huang}\ \emph {et~al.}(2024)\citenamefont {Huang},
  \citenamefont {Wang}, \citenamefont {Ye}, \citenamefont {Bao}, \citenamefont
  {Shangguan}, \citenamefont {Liao}, \citenamefont {Cao}, \citenamefont
  {Kajimoto}, \citenamefont {Ikeuchi}, \citenamefont {Deng}, \citenamefont
  {Smidman}, \citenamefont {Song}, \citenamefont {Yu}, \citenamefont {Li},\
  and\ \citenamefont {Wen}}]{Huang2024}%
  \BibitemOpen
  \bibfield  {author} {\bibinfo {author} {\bibfnamefont {Zhentao}\ \bibnamefont
  {Huang}}, \bibinfo {author} {\bibfnamefont {Wei}\ \bibnamefont {Wang}},
  \bibinfo {author} {\bibfnamefont {Huiqing}\ \bibnamefont {Ye}}, \bibinfo
  {author} {\bibfnamefont {Song}\ \bibnamefont {Bao}}, \bibinfo {author}
  {\bibfnamefont {Yanyan}\ \bibnamefont {Shangguan}}, \bibinfo {author}
  {\bibfnamefont {Junbo}\ \bibnamefont {Liao}}, \bibinfo {author}
  {\bibfnamefont {Saizheng}\ \bibnamefont {Cao}}, \bibinfo {author}
  {\bibfnamefont {Ryoichi}\ \bibnamefont {Kajimoto}}, \bibinfo {author}
  {\bibfnamefont {Kazuhiko}\ \bibnamefont {Ikeuchi}}, \bibinfo {author}
  {\bibfnamefont {Guochu}\ \bibnamefont {Deng}}, \bibinfo {author}
  {\bibfnamefont {Michael}\ \bibnamefont {Smidman}}, \bibinfo {author}
  {\bibfnamefont {Yu}~\bibnamefont {Song}}, \bibinfo {author} {\bibfnamefont
  {Shun-Li}\ \bibnamefont {Yu}}, \bibinfo {author} {\bibfnamefont {Jian-Xin}\
  \bibnamefont {Li}}, \ and\ \bibinfo {author} {\bibfnamefont {Jinsheng}\
  \bibnamefont {Wen}},\ }\bibfield  {title} {\enquote {\bibinfo {title}
  {{Microscopic origin of the spin-reorientation transition in the kagome
  topological magnet TbMn$_6$Sn$_6$}},}\ }\href {\doibase
  10.1103/physrevb.109.014434} {\bibfield  {journal} {\bibinfo  {journal}
  {Physical Review B}\ }\textbf {\bibinfo {volume} {109}},\ \bibinfo {pages}
  {014434} (\bibinfo {year} {2024})}\BibitemShut {NoStop}%
\bibitem [{\citenamefont {Weber}\ \emph {et~al.}(2022)\citenamefont {Weber},
  \citenamefont {Guennou}, \citenamefont {Evans}, \citenamefont {Toulouse},
  \citenamefont {Simonov}, \citenamefont {Kholina}, \citenamefont {Ma},
  \citenamefont {Ren}, \citenamefont {Cao}, \citenamefont {Carpenter},
  \citenamefont {Dkhil}, \citenamefont {Fiebig},\ and\ \citenamefont
  {Kreisel}}]{Weber2022}%
  \BibitemOpen
  \bibfield  {author} {\bibinfo {author} {\bibfnamefont {Mads~C.}\ \bibnamefont
  {Weber}}, \bibinfo {author} {\bibfnamefont {Mael}\ \bibnamefont {Guennou}},
  \bibinfo {author} {\bibfnamefont {Donald~M.}\ \bibnamefont {Evans}}, \bibinfo
  {author} {\bibfnamefont {Constance}\ \bibnamefont {Toulouse}}, \bibinfo
  {author} {\bibfnamefont {Arkadiy}\ \bibnamefont {Simonov}}, \bibinfo {author}
  {\bibfnamefont {Yevheniia}\ \bibnamefont {Kholina}}, \bibinfo {author}
  {\bibfnamefont {Xiaoxuan}\ \bibnamefont {Ma}}, \bibinfo {author}
  {\bibfnamefont {Wei}\ \bibnamefont {Ren}}, \bibinfo {author} {\bibfnamefont
  {Shixun}\ \bibnamefont {Cao}}, \bibinfo {author} {\bibfnamefont {Michael~A.}\
  \bibnamefont {Carpenter}}, \bibinfo {author} {\bibfnamefont {Brahim}\
  \bibnamefont {Dkhil}}, \bibinfo {author} {\bibfnamefont {Manfred}\
  \bibnamefont {Fiebig}}, \ and\ \bibinfo {author} {\bibfnamefont {Jens}\
  \bibnamefont {Kreisel}},\ }\bibfield  {title} {\enquote {\bibinfo {title}
  {{Emerging spin–phonon coupling through cross-talk of two magnetic
  sublattices}},}\ }\href {\doibase 10.1038/s41467-021-27267-8} {\bibfield
  {journal} {\bibinfo  {journal} {Nature Communications}\ }\textbf {\bibinfo
  {volume} {13}} (\bibinfo {year} {2022}),\
  10.1038/s41467-021-27267-8}\BibitemShut {NoStop}%
\bibitem [{\citenamefont {Levinson}\ \emph {et~al.}(1969)\citenamefont
  {Levinson}, \citenamefont {Luban},\ and\ \citenamefont
  {Shtrikman}}]{Levinson1969}%
  \BibitemOpen
  \bibfield  {author} {\bibinfo {author} {\bibfnamefont {Lionel~M.}\
  \bibnamefont {Levinson}}, \bibinfo {author} {\bibfnamefont {Marshall}\
  \bibnamefont {Luban}}, \ and\ \bibinfo {author} {\bibfnamefont
  {S.}~\bibnamefont {Shtrikman}},\ }\bibfield  {title} {\enquote {\bibinfo
  {title} {{Microscopic Model for Reorientation of the Easy Axis of
  Magnetization}},}\ }\href {\doibase 10.1103/physrev.187.715} {\bibfield
  {journal} {\bibinfo  {journal} {Physical Review}\ }\textbf {\bibinfo {volume}
  {187}},\ \bibinfo {pages} {715--722} (\bibinfo {year} {1969})}\BibitemShut
  {NoStop}%
\bibitem [{\citenamefont {Belov}\ \emph {et~al.}(1976)\citenamefont {Belov},
  \citenamefont {Zvezdin}, \citenamefont {Kadomtseva},\ and\ \citenamefont
  {Levitin}}]{Belov1976}%
  \BibitemOpen
  \bibfield  {author} {\bibinfo {author} {\bibfnamefont {Konstantin~P}\
  \bibnamefont {Belov}}, \bibinfo {author} {\bibfnamefont {Anatolii~K}\
  \bibnamefont {Zvezdin}}, \bibinfo {author} {\bibfnamefont {Antonina~M}\
  \bibnamefont {Kadomtseva}}, \ and\ \bibinfo {author} {\bibfnamefont {R~Z}\
  \bibnamefont {Levitin}},\ }\bibfield  {title} {\enquote {\bibinfo {title}
  {{Spin-reorientation transitions in rare-earth magnets}},}\ }\href {\doibase
  10.1070/pu1976v019n07abeh005274} {\bibfield  {journal} {\bibinfo  {journal}
  {Soviet Physics Uspekhi}\ }\textbf {\bibinfo {volume} {19}},\ \bibinfo
  {pages} {574--596} (\bibinfo {year} {1976})}\BibitemShut {NoStop}%
\bibitem [{\citenamefont {Yuan}\ \emph {et~al.}(2013)\citenamefont {Yuan},
  \citenamefont {Cao}, \citenamefont {Li}, \citenamefont {Qi}, \citenamefont
  {Cao}, \citenamefont {Zhang}, \citenamefont {DeLong},\ and\ \citenamefont
  {Cao}}]{Yuan2013}%
  \BibitemOpen
  \bibfield  {author} {\bibinfo {author} {\bibfnamefont {S.~J.}\ \bibnamefont
  {Yuan}}, \bibinfo {author} {\bibfnamefont {Y.~M.}\ \bibnamefont {Cao}},
  \bibinfo {author} {\bibfnamefont {L.}~\bibnamefont {Li}}, \bibinfo {author}
  {\bibfnamefont {T.~F.}\ \bibnamefont {Qi}}, \bibinfo {author} {\bibfnamefont
  {S.~X.}\ \bibnamefont {Cao}}, \bibinfo {author} {\bibfnamefont {J.~C.}\
  \bibnamefont {Zhang}}, \bibinfo {author} {\bibfnamefont {L.~E.}\ \bibnamefont
  {DeLong}}, \ and\ \bibinfo {author} {\bibfnamefont {G.}~\bibnamefont {Cao}},\
  }\bibfield  {title} {\enquote {\bibinfo {title} {{First-order spin
  reorientation transition and specific-heat anomaly in CeFeO3}},}\ }\href
  {\doibase 10.1063/1.4821516} {\bibfield  {journal} {\bibinfo  {journal}
  {Journal of Applied Physics}\ }\textbf {\bibinfo {volume} {114}} (\bibinfo
  {year} {2013}),\ 10.1063/1.4821516}\BibitemShut {NoStop}%
\bibitem [{\citenamefont {Song}\ \emph {et~al.}(2024)\citenamefont {Song},
  \citenamefont {Zhang}, \citenamefont {Li}, \citenamefont {Zhong},
  \citenamefont {Long}, \citenamefont {Wang}, \citenamefont {Xu}, \citenamefont
  {Zheng}, \citenamefont {Zhang}, \citenamefont {Huang}, \citenamefont {Zhang},
  \citenamefont {Xing},\ and\ \citenamefont {Chen}}]{Song2024}%
  \BibitemOpen
  \bibfield  {author} {\bibinfo {author} {\bibfnamefont {Yuzhu}\ \bibnamefont
  {Song}}, \bibinfo {author} {\bibfnamefont {Jimin}\ \bibnamefont {Zhang}},
  \bibinfo {author} {\bibfnamefont {Hengchao}\ \bibnamefont {Li}}, \bibinfo
  {author} {\bibfnamefont {Hong}\ \bibnamefont {Zhong}}, \bibinfo {author}
  {\bibfnamefont {Feixiang}\ \bibnamefont {Long}}, \bibinfo {author}
  {\bibfnamefont {Zhan}\ \bibnamefont {Wang}}, \bibinfo {author} {\bibfnamefont
  {Yuanji}\ \bibnamefont {Xu}}, \bibinfo {author} {\bibfnamefont {Xinqi}\
  \bibnamefont {Zheng}}, \bibinfo {author} {\bibfnamefont {Hu}~\bibnamefont
  {Zhang}}, \bibinfo {author} {\bibfnamefont {Qingzhen}\ \bibnamefont {Huang}},
  \bibinfo {author} {\bibfnamefont {Ying}\ \bibnamefont {Zhang}}, \bibinfo
  {author} {\bibfnamefont {Xianran}\ \bibnamefont {Xing}}, \ and\ \bibinfo
  {author} {\bibfnamefont {Jun}\ \bibnamefont {Chen}},\ }\bibfield  {title}
  {\enquote {\bibinfo {title} {{Unusual Magnetocaloric Effect Triggered by Spin
  Reorientation}},}\ }\href {\doibase 10.1002/aenm.202402527} {\bibfield
  {journal} {\bibinfo  {journal} {Advanced Energy Materials}\ } (\bibinfo
  {year} {2024}),\ 10.1002/aenm.202402527}\BibitemShut {NoStop}%
\bibitem [{\citenamefont {Law}\ \emph {et~al.}(2018)\citenamefont {Law},
  \citenamefont {Franco}, \citenamefont {Moreno-Ramírez}, \citenamefont
  {Conde}, \citenamefont {Karpenkov}, \citenamefont {Radulov}, \citenamefont
  {Skokov},\ and\ \citenamefont {Gutfleisch}}]{Law2018}%
  \BibitemOpen
  \bibfield  {author} {\bibinfo {author} {\bibfnamefont {Jia~Yan}\ \bibnamefont
  {Law}}, \bibinfo {author} {\bibfnamefont {Victorino}\ \bibnamefont {Franco}},
  \bibinfo {author} {\bibfnamefont {Luis~Miguel}\ \bibnamefont
  {Moreno-Ramírez}}, \bibinfo {author} {\bibfnamefont {Alejandro}\
  \bibnamefont {Conde}}, \bibinfo {author} {\bibfnamefont {Dmitriy~Y.}\
  \bibnamefont {Karpenkov}}, \bibinfo {author} {\bibfnamefont {Iliya}\
  \bibnamefont {Radulov}}, \bibinfo {author} {\bibfnamefont {Konstantin~P.}\
  \bibnamefont {Skokov}}, \ and\ \bibinfo {author} {\bibfnamefont {Oliver}\
  \bibnamefont {Gutfleisch}},\ }\bibfield  {title} {\enquote {\bibinfo {title}
  {{A quantitative criterion for determining the order of magnetic phase
  transitions using the magnetocaloric effect}},}\ }\href {\doibase
  10.1038/s41467-018-05111-w} {\bibfield  {journal} {\bibinfo  {journal}
  {Nature Communications}\ }\textbf {\bibinfo {volume} {9}} (\bibinfo {year}
  {2018}),\ 10.1038/s41467-018-05111-w}\BibitemShut {NoStop}%
\bibitem [{\citenamefont {Horner}\ and\ \citenamefont
  {Varma}(1968)}]{Horner1968}%
  \BibitemOpen
  \bibfield  {author} {\bibinfo {author} {\bibfnamefont {H.}~\bibnamefont
  {Horner}}\ and\ \bibinfo {author} {\bibfnamefont {C.~M.}\ \bibnamefont
  {Varma}},\ }\bibfield  {title} {\enquote {\bibinfo {title} {{Nature of
  Spin-Reorientation Transitions}},}\ }\href {\doibase
  10.1103/physrevlett.20.845} {\bibfield  {journal} {\bibinfo  {journal}
  {Physical Review Letters}\ }\textbf {\bibinfo {volume} {20}},\ \bibinfo
  {pages} {845--846} (\bibinfo {year} {1968})}\BibitemShut {NoStop}%
\bibitem [{\citenamefont {Moore}\ \emph {et~al.}(2022)\citenamefont {Moore},
  \citenamefont {Okamoto}, \citenamefont {Li}, \citenamefont {Meier},
  \citenamefont {Miao}, \citenamefont {Lee}, \citenamefont {Hashimoto},
  \citenamefont {Lu}, \citenamefont {Dagotto}, \citenamefont {McGuire},\ and\
  \citenamefont {Sales}}]{Moore2022}%
  \BibitemOpen
  \bibfield  {author} {\bibinfo {author} {\bibfnamefont {Robert~G.}\
  \bibnamefont {Moore}}, \bibinfo {author} {\bibfnamefont {Satoshi}\
  \bibnamefont {Okamoto}}, \bibinfo {author} {\bibfnamefont {Haoxiang}\
  \bibnamefont {Li}}, \bibinfo {author} {\bibfnamefont {William~R.}\
  \bibnamefont {Meier}}, \bibinfo {author} {\bibfnamefont {Hu}~\bibnamefont
  {Miao}}, \bibinfo {author} {\bibfnamefont {Ho~Nyung}\ \bibnamefont {Lee}},
  \bibinfo {author} {\bibfnamefont {Makoto}\ \bibnamefont {Hashimoto}},
  \bibinfo {author} {\bibfnamefont {Donghui}\ \bibnamefont {Lu}}, \bibinfo
  {author} {\bibfnamefont {Elbio}\ \bibnamefont {Dagotto}}, \bibinfo {author}
  {\bibfnamefont {Michael~A.}\ \bibnamefont {McGuire}}, \ and\ \bibinfo
  {author} {\bibfnamefont {Brian~C.}\ \bibnamefont {Sales}},\ }\bibfield
  {title} {\enquote {\bibinfo {title} {{Topological electronic structure
  evolution with symmetry-breaking spin reorientation in
  (Fe$_{1-x}$Co$_x$)Sn}},}\ }\href {\doibase 10.1103/physrevb.106.115141}
  {\bibfield  {journal} {\bibinfo  {journal} {Physical Review B}\ }\textbf
  {\bibinfo {volume} {106}},\ \bibinfo {pages} {115141} (\bibinfo {year}
  {2022})}\BibitemShut {NoStop}%
\bibitem [{\citenamefont {Karube}\ \emph {et~al.}(2022)\citenamefont {Karube},
  \citenamefont {Peng}, \citenamefont {Masell}, \citenamefont {Hemmida},
  \citenamefont {Krug~von Nidda}, \citenamefont {Kézsmárki}, \citenamefont
  {Yu}, \citenamefont {Tokura},\ and\ \citenamefont {Taguchi}}]{Karube2022}%
  \BibitemOpen
  \bibfield  {author} {\bibinfo {author} {\bibfnamefont {Kosuke}\ \bibnamefont
  {Karube}}, \bibinfo {author} {\bibfnamefont {Licong}\ \bibnamefont {Peng}},
  \bibinfo {author} {\bibfnamefont {Jan}\ \bibnamefont {Masell}}, \bibinfo
  {author} {\bibfnamefont {Mamoun}\ \bibnamefont {Hemmida}}, \bibinfo {author}
  {\bibfnamefont {Hans‐Albrecht}\ \bibnamefont {Krug~von Nidda}}, \bibinfo
  {author} {\bibfnamefont {István}\ \bibnamefont {Kézsmárki}}, \bibinfo
  {author} {\bibfnamefont {Xiuzhen}\ \bibnamefont {Yu}}, \bibinfo {author}
  {\bibfnamefont {Yoshinori}\ \bibnamefont {Tokura}}, \ and\ \bibinfo {author}
  {\bibfnamefont {Yasujiro}\ \bibnamefont {Taguchi}},\ }\bibfield  {title}
  {\enquote {\bibinfo {title} {{Doping Control of Magnetic Anisotropy for
  Stable Antiskyrmion Formation in Schreibersite (Fe,Ni)$_3$P with S4
  symmetry}},}\ }\href {\doibase 10.1002/adma.202108770} {\bibfield  {journal}
  {\bibinfo  {journal} {Advanced Materials}\ }\textbf {\bibinfo {volume} {34}}
  (\bibinfo {year} {2022}),\ 10.1002/adma.202108770}\BibitemShut {NoStop}%
\bibitem [{\citenamefont {Masuda}\ \emph {et~al.}(2020)\citenamefont {Masuda},
  \citenamefont {Sakai}, \citenamefont {Takahashi}, \citenamefont {Yamasaki},
  \citenamefont {Nakao}, \citenamefont {Moyoshi}, \citenamefont {Nakao},
  \citenamefont {Murakami}, \citenamefont {Arima},\ and\ \citenamefont
  {Ishiwata}}]{Masuda2020}%
  \BibitemOpen
  \bibfield  {author} {\bibinfo {author} {\bibfnamefont {H.}~\bibnamefont
  {Masuda}}, \bibinfo {author} {\bibfnamefont {H.}~\bibnamefont {Sakai}},
  \bibinfo {author} {\bibfnamefont {H.}~\bibnamefont {Takahashi}}, \bibinfo
  {author} {\bibfnamefont {Y.}~\bibnamefont {Yamasaki}}, \bibinfo {author}
  {\bibfnamefont {A.}~\bibnamefont {Nakao}}, \bibinfo {author} {\bibfnamefont
  {T.}~\bibnamefont {Moyoshi}}, \bibinfo {author} {\bibfnamefont
  {H.}~\bibnamefont {Nakao}}, \bibinfo {author} {\bibfnamefont
  {Y.}~\bibnamefont {Murakami}}, \bibinfo {author} {\bibfnamefont
  {T.}~\bibnamefont {Arima}}, \ and\ \bibinfo {author} {\bibfnamefont
  {S.}~\bibnamefont {Ishiwata}},\ }\bibfield  {title} {\enquote {\bibinfo
  {title} {{Field-induced spin reorientation in the antiferromagnetic Dirac
  material EuMnBi$_2$ revealed by neutron and resonant x-ray diffraction}},}\
  }\href {\doibase 10.1103/physrevb.101.174411} {\bibfield  {journal} {\bibinfo
   {journal} {Physical Review B}\ }\textbf {\bibinfo {volume} {101}},\ \bibinfo
  {pages} {174411} (\bibinfo {year} {2020})}\BibitemShut {NoStop}%
\bibitem [{\citenamefont {Yamamoto}\ \emph {et~al.}(2021)\citenamefont
  {Yamamoto}, \citenamefont {Suwa}, \citenamefont {Kihara}, \citenamefont
  {Nomura}, \citenamefont {Kotani}, \citenamefont {Nakamura}, \citenamefont
  {Skourski}, \citenamefont {Zherlitsyn}, \citenamefont {Prodan}, \citenamefont
  {Tsurkan}, \citenamefont {Nojiri}, \citenamefont {Loidl},\ and\ \citenamefont
  {Wosnitza}}]{Yamamoto2021}%
  \BibitemOpen
  \bibfield  {author} {\bibinfo {author} {\bibfnamefont {Sh.}\ \bibnamefont
  {Yamamoto}}, \bibinfo {author} {\bibfnamefont {H.}~\bibnamefont {Suwa}},
  \bibinfo {author} {\bibfnamefont {T.}~\bibnamefont {Kihara}}, \bibinfo
  {author} {\bibfnamefont {T.}~\bibnamefont {Nomura}}, \bibinfo {author}
  {\bibfnamefont {Y.}~\bibnamefont {Kotani}}, \bibinfo {author} {\bibfnamefont
  {T.}~\bibnamefont {Nakamura}}, \bibinfo {author} {\bibfnamefont
  {Y.}~\bibnamefont {Skourski}}, \bibinfo {author} {\bibfnamefont
  {S.}~\bibnamefont {Zherlitsyn}}, \bibinfo {author} {\bibfnamefont
  {L.}~\bibnamefont {Prodan}}, \bibinfo {author} {\bibfnamefont
  {V.}~\bibnamefont {Tsurkan}}, \bibinfo {author} {\bibfnamefont
  {H.}~\bibnamefont {Nojiri}}, \bibinfo {author} {\bibfnamefont
  {A.}~\bibnamefont {Loidl}}, \ and\ \bibinfo {author} {\bibfnamefont
  {J.}~\bibnamefont {Wosnitza}},\ }\bibfield  {title} {\enquote {\bibinfo
  {title} {{Element-specific field-induced spin reorientation and tetracritical
  point in MnCr$_2$S$_4$}},}\ }\href {\doibase 10.1103/physrevb.103.l020408}
  {\bibfield  {journal} {\bibinfo  {journal} {Physical Review B}\ }\textbf
  {\bibinfo {volume} {103}},\ \bibinfo {pages} {l020408} (\bibinfo {year}
  {2021})}\BibitemShut {NoStop}%
\bibitem [{\citenamefont {Prodan}\ \emph {et~al.}(2021)\citenamefont {Prodan},
  \citenamefont {Yasin}, \citenamefont {Jesche}, \citenamefont {Deisenhofer},
  \citenamefont {von Nidda}, \citenamefont {Mayr}, \citenamefont {Zherlitsyn},
  \citenamefont {Wosnitza}, \citenamefont {Loidl},\ and\ \citenamefont
  {Tsurkan}}]{Prodan2021}%
  \BibitemOpen
  \bibfield  {author} {\bibinfo {author} {\bibfnamefont {L.}~\bibnamefont
  {Prodan}}, \bibinfo {author} {\bibfnamefont {S.}~\bibnamefont {Yasin}},
  \bibinfo {author} {\bibfnamefont {A.}~\bibnamefont {Jesche}}, \bibinfo
  {author} {\bibfnamefont {J.}~\bibnamefont {Deisenhofer}}, \bibinfo {author}
  {\bibfnamefont {H.-A.~Krug}\ \bibnamefont {von Nidda}}, \bibinfo {author}
  {\bibfnamefont {F.}~\bibnamefont {Mayr}}, \bibinfo {author} {\bibfnamefont
  {S.}~\bibnamefont {Zherlitsyn}}, \bibinfo {author} {\bibfnamefont
  {J.}~\bibnamefont {Wosnitza}}, \bibinfo {author} {\bibfnamefont
  {A.}~\bibnamefont {Loidl}}, \ and\ \bibinfo {author} {\bibfnamefont
  {V.}~\bibnamefont {Tsurkan}},\ }\bibfield  {title} {\enquote {\bibinfo
  {title} {{Unusual field-induced spin reorientation in FeCr$_2$S$_4$ : Field
  tuning of the Jahn-Teller state}},}\ }\href {\doibase
  10.1103/physrevb.104.l020410} {\bibfield  {journal} {\bibinfo  {journal}
  {Physical Review B}\ }\textbf {\bibinfo {volume} {104}},\ \bibinfo {pages}
  {l020410} (\bibinfo {year} {2021})}\BibitemShut {NoStop}%
\bibitem [{\citenamefont {Lin}\ \emph {et~al.}(2018{\natexlab{a}})\citenamefont
  {Lin}, \citenamefont {Lohmann}, \citenamefont {Ali}, \citenamefont {Tang},
  \citenamefont {Li}, \citenamefont {Xing}, \citenamefont {Zhong},
  \citenamefont {Jia}, \citenamefont {Han}, \citenamefont {Coh}, \citenamefont
  {Beyermann},\ and\ \citenamefont {Shi}}]{Lin2018}%
  \BibitemOpen
  \bibfield  {author} {\bibinfo {author} {\bibfnamefont {Zhisheng}\
  \bibnamefont {Lin}}, \bibinfo {author} {\bibfnamefont {Mark}\ \bibnamefont
  {Lohmann}}, \bibinfo {author} {\bibfnamefont {Zulfikhar~A.}\ \bibnamefont
  {Ali}}, \bibinfo {author} {\bibfnamefont {Chi}\ \bibnamefont {Tang}},
  \bibinfo {author} {\bibfnamefont {Junxue}\ \bibnamefont {Li}}, \bibinfo
  {author} {\bibfnamefont {Wenyu}\ \bibnamefont {Xing}}, \bibinfo {author}
  {\bibfnamefont {Jiangnan}\ \bibnamefont {Zhong}}, \bibinfo {author}
  {\bibfnamefont {Shuang}\ \bibnamefont {Jia}}, \bibinfo {author}
  {\bibfnamefont {Wei}\ \bibnamefont {Han}}, \bibinfo {author} {\bibfnamefont
  {Sinisa}\ \bibnamefont {Coh}}, \bibinfo {author} {\bibfnamefont {Ward}\
  \bibnamefont {Beyermann}}, \ and\ \bibinfo {author} {\bibfnamefont {Jing}\
  \bibnamefont {Shi}},\ }\bibfield  {title} {\enquote {\bibinfo {title}
  {{Pressure-induced spin reorientation transition in layered ferromagnetic
  insulator Cr$_2$Ge$_2$Te$_6$}},}\ }\href {\doibase
  10.1103/physrevmaterials.2.051004} {\bibfield  {journal} {\bibinfo  {journal}
  {Physical Review Materials}\ }\textbf {\bibinfo {volume} {2}},\ \bibinfo
  {pages} {051004} (\bibinfo {year} {2018}{\natexlab{a}})}\BibitemShut
  {NoStop}%
\bibitem [{\citenamefont {Skorobogatov}\ \emph {et~al.}(2023)\citenamefont
  {Skorobogatov}, \citenamefont {Wu}, \citenamefont {Xie}, \citenamefont
  {Shaykhutdinov}, \citenamefont {Pomjakushina}, \citenamefont {Podlesnyak},\
  and\ \citenamefont {Nikitin}}]{Skorobogatov2023}%
  \BibitemOpen
  \bibfield  {author} {\bibinfo {author} {\bibfnamefont {S.~A.}\ \bibnamefont
  {Skorobogatov}}, \bibinfo {author} {\bibfnamefont {L.~S.}\ \bibnamefont
  {Wu}}, \bibinfo {author} {\bibfnamefont {T.}~\bibnamefont {Xie}}, \bibinfo
  {author} {\bibfnamefont {K.~A.}\ \bibnamefont {Shaykhutdinov}}, \bibinfo
  {author} {\bibfnamefont {E.~V.}\ \bibnamefont {Pomjakushina}}, \bibinfo
  {author} {\bibfnamefont {A.}~\bibnamefont {Podlesnyak}}, \ and\ \bibinfo
  {author} {\bibfnamefont {S.~E.}\ \bibnamefont {Nikitin}},\ }\bibfield
  {title} {\enquote {\bibinfo {title} {{Pressure control of the spin
  reorientation transition in the rare-earth orthoferrite YbFeO$_3$}},}\ }\href
  {\doibase 10.1103/physrevb.108.054432} {\bibfield  {journal} {\bibinfo
  {journal} {Physical Review B}\ }\textbf {\bibinfo {volume} {108}},\ \bibinfo
  {pages} {054432} (\bibinfo {year} {2023})}\BibitemShut {NoStop}%
\bibitem [{\citenamefont {Bordel}\ \emph {et~al.}(2012)\citenamefont {Bordel},
  \citenamefont {Juraszek}, \citenamefont {Cooke}, \citenamefont
  {Baldasseroni}, \citenamefont {Mankovsky}, \citenamefont {Minár},
  \citenamefont {Ebert}, \citenamefont {Moyerman}, \citenamefont {Fullerton},\
  and\ \citenamefont {Hellman}}]{Bordel2012}%
  \BibitemOpen
  \bibfield  {author} {\bibinfo {author} {\bibfnamefont {C.}~\bibnamefont
  {Bordel}}, \bibinfo {author} {\bibfnamefont {J.}~\bibnamefont {Juraszek}},
  \bibinfo {author} {\bibfnamefont {David~W.}\ \bibnamefont {Cooke}}, \bibinfo
  {author} {\bibfnamefont {C.}~\bibnamefont {Baldasseroni}}, \bibinfo {author}
  {\bibfnamefont {S.}~\bibnamefont {Mankovsky}}, \bibinfo {author}
  {\bibfnamefont {J.}~\bibnamefont {Minár}}, \bibinfo {author} {\bibfnamefont
  {H.}~\bibnamefont {Ebert}}, \bibinfo {author} {\bibfnamefont
  {S.}~\bibnamefont {Moyerman}}, \bibinfo {author} {\bibfnamefont {E.~E.}\
  \bibnamefont {Fullerton}}, \ and\ \bibinfo {author} {\bibfnamefont
  {F.}~\bibnamefont {Hellman}},\ }\bibfield  {title} {\enquote {\bibinfo
  {title} {{Fe Spin Reorientation across the Metamagnetic Transition in
  Strained FeRh Thin Films}},}\ }\href {\doibase
  10.1103/physrevlett.109.117201} {\bibfield  {journal} {\bibinfo  {journal}
  {Physical Review Letters}\ }\textbf {\bibinfo {volume} {109}},\ \bibinfo
  {pages} {117201} (\bibinfo {year} {2012})}\BibitemShut {NoStop}%
\bibitem [{\citenamefont {Kong}\ \emph {et~al.}(2024)\citenamefont {Kong},
  \citenamefont {Kovács}, \citenamefont {Charilaou}, \citenamefont
  {Altthaler}, \citenamefont {Prodan}, \citenamefont {Tsuran}, \citenamefont
  {Meier}, \citenamefont {Han}, \citenamefont {Kezsmarki},\ and\ \citenamefont
  {Dunin-Borkowski}}]{Kong2024}%
  \BibitemOpen
  \bibfield  {author} {\bibinfo {author} {\bibfnamefont {D.}~\bibnamefont
  {Kong}}, \bibinfo {author} {\bibfnamefont {A.}~\bibnamefont {Kovács}},
  \bibinfo {author} {\bibfnamefont {M.}~\bibnamefont {Charilaou}}, \bibinfo
  {author} {\bibfnamefont {M.}~\bibnamefont {Altthaler}}, \bibinfo {author}
  {\bibfnamefont {L.}~\bibnamefont {Prodan}}, \bibinfo {author} {\bibfnamefont
  {V.}~\bibnamefont {Tsuran}}, \bibinfo {author} {\bibfnamefont
  {D.}~\bibnamefont {Meier}}, \bibinfo {author} {\bibfnamefont
  {X.}~\bibnamefont {Han}}, \bibinfo {author} {\bibfnamefont {I}~\bibnamefont
  {Kezsmarki}}, \ and\ \bibinfo {author} {\bibfnamefont {R.~E.}\ \bibnamefont
  {Dunin-Borkowski}},\ }\href {\doibase 10.48550/ARXIV.2412.12684} {\enquote
  {\bibinfo {title} {{Strain engineering of magnetic anisotropy in the kagome
  magnet Fe$_3$Sn$_2$}},}\ } (\bibinfo {year} {2024})\BibitemShut {NoStop}%
\bibitem [{\citenamefont {Kimel}\ \emph {et~al.}(2004)\citenamefont {Kimel},
  \citenamefont {Kirilyuk}, \citenamefont {Tsvetkov}, \citenamefont {Pisarev},\
  and\ \citenamefont {Rasing}}]{Kimel2004}%
  \BibitemOpen
  \bibfield  {author} {\bibinfo {author} {\bibfnamefont {A.~V.}\ \bibnamefont
  {Kimel}}, \bibinfo {author} {\bibfnamefont {A.}~\bibnamefont {Kirilyuk}},
  \bibinfo {author} {\bibfnamefont {A.}~\bibnamefont {Tsvetkov}}, \bibinfo
  {author} {\bibfnamefont {R.~V.}\ \bibnamefont {Pisarev}}, \ and\ \bibinfo
  {author} {\bibfnamefont {Th.}\ \bibnamefont {Rasing}},\ }\bibfield  {title}
  {\enquote {\bibinfo {title} {{Laser-induced ultrafast spin reorientation in
  the antiferromagnet TmFeO$_3$}},}\ }\href {\doibase 10.1038/nature02659}
  {\bibfield  {journal} {\bibinfo  {journal} {Nature}\ }\textbf {\bibinfo
  {volume} {429}},\ \bibinfo {pages} {850--853} (\bibinfo {year}
  {2004})}\BibitemShut {NoStop}%
\bibitem [{\citenamefont {Kézsmárki}\ \emph {et~al.}(2015)\citenamefont
  {Kézsmárki}, \citenamefont {Bordács}, \citenamefont {Milde}, \citenamefont
  {Neuber}, \citenamefont {Eng}, \citenamefont {White}, \citenamefont
  {Rønnow}, \citenamefont {Dewhurst}, \citenamefont {Mochizuki}, \citenamefont
  {Yanai}, \citenamefont {Nakamura}, \citenamefont {Ehlers}, \citenamefont
  {Tsurkan},\ and\ \citenamefont {Loidl}}]{Kezsmarki2015}%
  \BibitemOpen
  \bibfield  {author} {\bibinfo {author} {\bibfnamefont {I.}~\bibnamefont
  {Kézsmárki}}, \bibinfo {author} {\bibfnamefont {S.}~\bibnamefont
  {Bordács}}, \bibinfo {author} {\bibfnamefont {P.}~\bibnamefont {Milde}},
  \bibinfo {author} {\bibfnamefont {E.}~\bibnamefont {Neuber}}, \bibinfo
  {author} {\bibfnamefont {L. M.}\ \bibnamefont {Eng}}, \bibinfo {author}
  {\bibfnamefont {J. S.}\ \bibnamefont {White}}, \bibinfo {author}
  {\bibfnamefont {H. M.}\ \bibnamefont {Rønnow}}, \bibinfo {author}
  {\bibfnamefont {C. D.}\ \bibnamefont {Dewhurst}}, \bibinfo {author}
  {\bibfnamefont {M.}~\bibnamefont {Mochizuki}}, \bibinfo {author}
  {\bibfnamefont {K.}~\bibnamefont {Yanai}}, \bibinfo {author} {\bibfnamefont
  {H.}~\bibnamefont {Nakamura}}, \bibinfo {author} {\bibfnamefont
  {D.}~\bibnamefont {Ehlers}}, \bibinfo {author} {\bibfnamefont
  {V.}~\bibnamefont {Tsurkan}}, \ and\ \bibinfo {author} {\bibfnamefont
  {A.}~\bibnamefont {Loidl}},\ }\bibfield  {title} {\enquote {\bibinfo {title}
  {{Néel-type skyrmion lattice with confined orientation in the polar magnetic
  semiconductor GaV$_4$S$_8$}},}\ }\href {\doibase 10.1038/nmat4402}
  {\bibfield  {journal} {\bibinfo  {journal} {Nature Materials}\ }\textbf
  {\bibinfo {volume} {14}},\ \bibinfo {pages} {1116--1122} (\bibinfo {year}
  {2015})}\BibitemShut {NoStop}%
\bibitem [{\citenamefont {Ehlers}\ \emph {et~al.}(2016)\citenamefont {Ehlers},
  \citenamefont {Stasinopoulos}, \citenamefont {Tsurkan}, \citenamefont
  {Krug~von Nidda}, \citenamefont {Fehér}, \citenamefont {Leonov},
  \citenamefont {Kézsmárki}, \citenamefont {Grundler},\ and\ \citenamefont
  {Loidl}}]{Ehlers2016}%
  \BibitemOpen
  \bibfield  {author} {\bibinfo {author} {\bibfnamefont {D.}~\bibnamefont
  {Ehlers}}, \bibinfo {author} {\bibfnamefont {I.}~\bibnamefont
  {Stasinopoulos}}, \bibinfo {author} {\bibfnamefont {V.}~\bibnamefont
  {Tsurkan}}, \bibinfo {author} {\bibfnamefont {H.-A.}\ \bibnamefont {Krug~von
  Nidda}}, \bibinfo {author} {\bibfnamefont {T.}~\bibnamefont {Fehér}},
  \bibinfo {author} {\bibfnamefont {A.}~\bibnamefont {Leonov}}, \bibinfo
  {author} {\bibfnamefont {I.}~\bibnamefont {Kézsmárki}}, \bibinfo {author}
  {\bibfnamefont {D.}~\bibnamefont {Grundler}}, \ and\ \bibinfo {author}
  {\bibfnamefont {A.}~\bibnamefont {Loidl}},\ }\bibfield  {title} {\enquote
  {\bibinfo {title} {{Skyrmion dynamics under uniaxial anisotropy}},}\ }\href
  {\doibase 10.1103/physrevb.94.014406} {\bibfield  {journal} {\bibinfo
  {journal} {Physical Review B}\ }\textbf {\bibinfo {volume} {94}},\ \bibinfo
  {pages} {014406} (\bibinfo {year} {2016})}\BibitemShut {NoStop}%
\bibitem [{\citenamefont {Bordács}\ \emph {et~al.}(2017)\citenamefont
  {Bordács}, \citenamefont {Butykai}, \citenamefont {Szigeti}, \citenamefont
  {White}, \citenamefont {Cubitt}, \citenamefont {Leonov}, \citenamefont
  {Widmann}, \citenamefont {Ehlers}, \citenamefont {von Nidda}, \citenamefont
  {Tsurkan}, \citenamefont {Loidl},\ and\ \citenamefont
  {Kézsmárki}}]{Bordacs2017}%
  \BibitemOpen
  \bibfield  {author} {\bibinfo {author} {\bibfnamefont {S.}~\bibnamefont
  {Bordács}}, \bibinfo {author} {\bibfnamefont {A.}~\bibnamefont {Butykai}},
  \bibinfo {author} {\bibfnamefont {B.~G.}\ \bibnamefont {Szigeti}}, \bibinfo
  {author} {\bibfnamefont {J.~S.}\ \bibnamefont {White}}, \bibinfo {author}
  {\bibfnamefont {R.}~\bibnamefont {Cubitt}}, \bibinfo {author} {\bibfnamefont
  {A.~O.}\ \bibnamefont {Leonov}}, \bibinfo {author} {\bibfnamefont
  {S.}~\bibnamefont {Widmann}}, \bibinfo {author} {\bibfnamefont
  {D.}~\bibnamefont {Ehlers}}, \bibinfo {author} {\bibfnamefont {H.-A.~Krug}\
  \bibnamefont {von Nidda}}, \bibinfo {author} {\bibfnamefont {V.}~\bibnamefont
  {Tsurkan}}, \bibinfo {author} {\bibfnamefont {A.}~\bibnamefont {Loidl}}, \
  and\ \bibinfo {author} {\bibfnamefont {I.}~\bibnamefont {Kézsmárki}},\
  }\bibfield  {title} {\enquote {\bibinfo {title} {{Equilibrium Skyrmion
  Lattice Ground State in a Polar Easy-plane Magnet}},}\ }\href {\doibase
  10.1038/s41598-017-07996-x} {\bibfield  {journal} {\bibinfo  {journal}
  {Scientific Reports}\ }\textbf {\bibinfo {volume} {7}} (\bibinfo {year}
  {2017}),\ 10.1038/s41598-017-07996-x}\BibitemShut {NoStop}%
\bibitem [{\citenamefont {Leonov}\ and\ \citenamefont
  {Kézsmárki}(2017)}]{Leonov2017}%
  \BibitemOpen
  \bibfield  {author} {\bibinfo {author} {\bibfnamefont {A.~O.}\ \bibnamefont
  {Leonov}}\ and\ \bibinfo {author} {\bibfnamefont {I.}~\bibnamefont
  {Kézsmárki}},\ }\bibfield  {title} {\enquote {\bibinfo {title} {{Skyrmion
  robustness in noncentrosymmetric magnets with axial symmetry: The role of
  anisotropy and tilted magnetic fields}},}\ }\href {\doibase
  10.1103/physrevb.96.214413} {\bibfield  {journal} {\bibinfo  {journal}
  {Physical Review B}\ }\textbf {\bibinfo {volume} {96}},\ \bibinfo {pages}
  {214413} (\bibinfo {year} {2017})}\BibitemShut {NoStop}%
\bibitem [{\citenamefont {Preißinger}\ \emph {et~al.}(2021)\citenamefont
  {Preißinger}, \citenamefont {Karube}, \citenamefont {Ehlers}, \citenamefont
  {Szigeti}, \citenamefont {Krug~von Nidda}, \citenamefont {White},
  \citenamefont {Ukleev}, \citenamefont {Rønnow}, \citenamefont {Tokunaga},
  \citenamefont {Kikkawa}, \citenamefont {Tokura}, \citenamefont {Taguchi},\
  and\ \citenamefont {Kézsmárki}}]{Preissinger2021}%
  \BibitemOpen
  \bibfield  {author} {\bibinfo {author} {\bibfnamefont {M.}~\bibnamefont
  {Preißinger}}, \bibinfo {author} {\bibfnamefont {K.}~\bibnamefont {Karube}},
  \bibinfo {author} {\bibfnamefont {D.}~\bibnamefont {Ehlers}}, \bibinfo
  {author} {\bibfnamefont {B.}~\bibnamefont {Szigeti}}, \bibinfo {author}
  {\bibfnamefont {H.-A.}\ \bibnamefont {Krug~von Nidda}}, \bibinfo {author}
  {\bibfnamefont {J.~S.}\ \bibnamefont {White}}, \bibinfo {author}
  {\bibfnamefont {V.}~\bibnamefont {Ukleev}}, \bibinfo {author} {\bibfnamefont
  {H.~M.}\ \bibnamefont {Rønnow}}, \bibinfo {author} {\bibfnamefont
  {Y.}~\bibnamefont {Tokunaga}}, \bibinfo {author} {\bibfnamefont
  {A.}~\bibnamefont {Kikkawa}}, \bibinfo {author} {\bibfnamefont
  {Y.}~\bibnamefont {Tokura}}, \bibinfo {author} {\bibfnamefont
  {Y.}~\bibnamefont {Taguchi}}, \ and\ \bibinfo {author} {\bibfnamefont
  {I.}~\bibnamefont {Kézsmárki}},\ }\bibfield  {title} {\enquote {\bibinfo
  {title} {{Vital role of magnetocrystalline anisotropy in cubic chiral
  skyrmion hosts}},}\ }\href {\doibase 10.1038/s41535-021-00365-y} {\bibfield
  {journal} {\bibinfo  {journal} {npj Quantum Materials}\ }\textbf {\bibinfo
  {volume} {6}} (\bibinfo {year} {2021}),\
  10.1038/s41535-021-00365-y}\BibitemShut {NoStop}%
\bibitem [{\citenamefont {Meshcheriakova}\ \emph {et~al.}(2014)\citenamefont
  {Meshcheriakova}, \citenamefont {Chadov}, \citenamefont {Nayak},
  \citenamefont {Rößler}, \citenamefont {Kübler}, \citenamefont {André},
  \citenamefont {Tsirlin}, \citenamefont {Kiss}, \citenamefont {Hausdorf},
  \citenamefont {Kalache}, \citenamefont {Schnelle}, \citenamefont {Nicklas},\
  and\ \citenamefont {Felser}}]{Meshcheriakova2014}%
  \BibitemOpen
  \bibfield  {author} {\bibinfo {author} {\bibfnamefont {O.}~\bibnamefont
  {Meshcheriakova}}, \bibinfo {author} {\bibfnamefont {S.}~\bibnamefont
  {Chadov}}, \bibinfo {author} {\bibfnamefont {A.~K.}\ \bibnamefont {Nayak}},
  \bibinfo {author} {\bibfnamefont {U.~K.}\ \bibnamefont {Rößler}}, \bibinfo
  {author} {\bibfnamefont {J.}~\bibnamefont {Kübler}}, \bibinfo {author}
  {\bibfnamefont {G.}~\bibnamefont {André}}, \bibinfo {author} {\bibfnamefont
  {A.~A.}\ \bibnamefont {Tsirlin}}, \bibinfo {author} {\bibfnamefont
  {J.}~\bibnamefont {Kiss}}, \bibinfo {author} {\bibfnamefont {S.}~\bibnamefont
  {Hausdorf}}, \bibinfo {author} {\bibfnamefont {A.}~\bibnamefont {Kalache}},
  \bibinfo {author} {\bibfnamefont {W.}~\bibnamefont {Schnelle}}, \bibinfo
  {author} {\bibfnamefont {M.}~\bibnamefont {Nicklas}}, \ and\ \bibinfo
  {author} {\bibfnamefont {C.}~\bibnamefont {Felser}},\ }\bibfield  {title}
  {\enquote {\bibinfo {title} {{Large Noncollinearity and Spin Reorientation in
  the Novel Mn\textsubscript{2}RhSn Heusler Magnet}},}\ }\href {\doibase
  10.1103/PhysRevLett.113.087203} {\bibfield  {journal} {\bibinfo  {journal}
  {Physical Review Letters}\ }\textbf {\bibinfo {volume} {113}},\ \bibinfo
  {pages} {087203} (\bibinfo {year} {2014})}\BibitemShut {NoStop}%
\bibitem [{\citenamefont {Sukhanov}\ \emph {et~al.}(2020)\citenamefont
  {Sukhanov}, \citenamefont {Cespedes}, \citenamefont {Vir}, \citenamefont
  {Cameron}, \citenamefont {Heinemann}, \citenamefont {Martin}, \citenamefont
  {Chaboussant}, \citenamefont {Kumar}, \citenamefont {Milde}, \citenamefont
  {Eng}, \citenamefont {Felser},\ and\ \citenamefont {Inosov}}]{Sukhanov2020}%
  \BibitemOpen
  \bibfield  {author} {\bibinfo {author} {\bibfnamefont {A.~S.}\ \bibnamefont
  {Sukhanov}}, \bibinfo {author} {\bibfnamefont {B.~E.~Zuniga}\ \bibnamefont
  {Cespedes}}, \bibinfo {author} {\bibfnamefont {P.}~\bibnamefont {Vir}},
  \bibinfo {author} {\bibfnamefont {A.~S.}\ \bibnamefont {Cameron}}, \bibinfo
  {author} {\bibfnamefont {A.}~\bibnamefont {Heinemann}}, \bibinfo {author}
  {\bibfnamefont {N.}~\bibnamefont {Martin}}, \bibinfo {author} {\bibfnamefont
  {G.}~\bibnamefont {Chaboussant}}, \bibinfo {author} {\bibfnamefont
  {V.}~\bibnamefont {Kumar}}, \bibinfo {author} {\bibfnamefont
  {P.}~\bibnamefont {Milde}}, \bibinfo {author} {\bibfnamefont {L.~M.}\
  \bibnamefont {Eng}}, \bibinfo {author} {\bibfnamefont {C.}~\bibnamefont
  {Felser}}, \ and\ \bibinfo {author} {\bibfnamefont {D.~S.}\ \bibnamefont
  {Inosov}},\ }\bibfield  {title} {\enquote {\bibinfo {title} {{Anisotropic
  fractal magnetic domain pattern in bulk Mn$_{1.4}$PtSn}},}\ }\href {\doibase
  10.1103/physrevb.102.174447} {\bibfield  {journal} {\bibinfo  {journal}
  {Physical Review B}\ }\textbf {\bibinfo {volume} {102}},\ \bibinfo {pages}
  {174447} (\bibinfo {year} {2020})}\BibitemShut {NoStop}%
\bibitem [{\citenamefont {Xiao}\ \emph {et~al.}(2020)\citenamefont {Xiao},
  \citenamefont {Morvan}, \citenamefont {He}, \citenamefont {Wang},
  \citenamefont {Luo}, \citenamefont {Jiao}, \citenamefont {Xia}, \citenamefont
  {Zhao},\ and\ \citenamefont {Liu}}]{Xiao2020}%
  \BibitemOpen
  \bibfield  {author} {\bibinfo {author} {\bibfnamefont {Y.}~\bibnamefont
  {Xiao}}, \bibinfo {author} {\bibfnamefont {F.~J.}\ \bibnamefont {Morvan}},
  \bibinfo {author} {\bibfnamefont {A.~N.}\ \bibnamefont {He}}, \bibinfo
  {author} {\bibfnamefont {M.~K.}\ \bibnamefont {Wang}}, \bibinfo {author}
  {\bibfnamefont {H.~B.}\ \bibnamefont {Luo}}, \bibinfo {author} {\bibfnamefont
  {R.~B.}\ \bibnamefont {Jiao}}, \bibinfo {author} {\bibfnamefont {W.~X.}\
  \bibnamefont {Xia}}, \bibinfo {author} {\bibfnamefont {G.~P.}\ \bibnamefont
  {Zhao}}, \ and\ \bibinfo {author} {\bibfnamefont {J.~P.}\ \bibnamefont
  {Liu}},\ }\bibfield  {title} {\enquote {\bibinfo {title} {{Spin-reorientation
  transition induced magnetic skyrmion in Nd$_2$Fe$_{14}$B magnet}},}\ }\href
  {\doibase 10.1063/5.0022270} {\bibfield  {journal} {\bibinfo  {journal}
  {Applied Physics Letters}\ }\textbf {\bibinfo {volume} {117}} (\bibinfo
  {year} {2020}),\ 10.1063/5.0022270}\BibitemShut {NoStop}%
\bibitem [{\citenamefont {Bera}\ \emph {et~al.}(2023)\citenamefont {Bera},
  \citenamefont {Pradhan}, \citenamefont {Khan}, \citenamefont {Pal},
  \citenamefont {Pal}, \citenamefont {Kalimuddin}, \citenamefont {Bera},
  \citenamefont {Das}, \citenamefont {Pal},\ and\ \citenamefont
  {Mondal}}]{Bera2023}%
  \BibitemOpen
  \bibfield  {author} {\bibinfo {author} {\bibfnamefont {Satyabrata}\
  \bibnamefont {Bera}}, \bibinfo {author} {\bibfnamefont {Suman~Kalyan}\
  \bibnamefont {Pradhan}}, \bibinfo {author} {\bibfnamefont {Md~Salman}\
  \bibnamefont {Khan}}, \bibinfo {author} {\bibfnamefont {Riju}\ \bibnamefont
  {Pal}}, \bibinfo {author} {\bibfnamefont {Buddhadeb}\ \bibnamefont {Pal}},
  \bibinfo {author} {\bibfnamefont {Sk}~\bibnamefont {Kalimuddin}}, \bibinfo
  {author} {\bibfnamefont {Arnab}\ \bibnamefont {Bera}}, \bibinfo {author}
  {\bibfnamefont {Biswajit}\ \bibnamefont {Das}}, \bibinfo {author}
  {\bibfnamefont {Atindra~Nath}\ \bibnamefont {Pal}}, \ and\ \bibinfo {author}
  {\bibfnamefont {Mintu}\ \bibnamefont {Mondal}},\ }\bibfield  {title}
  {\enquote {\bibinfo {title} {{Unravelling the nature of spin reorientation
  transition in quasi-2D vdW magnetic material, Fe$_4$GeTe$_2$}},}\ }\href
  {\doibase 10.1016/j.jmmm.2022.170257} {\bibfield  {journal} {\bibinfo
  {journal} {Journal of Magnetism and Magnetic Materials}\ }\textbf {\bibinfo
  {volume} {565}},\ \bibinfo {pages} {170257} (\bibinfo {year}
  {2023})}\BibitemShut {NoStop}%
\bibitem [{\citenamefont {Wang}\ \emph
  {et~al.}(2024{\natexlab{a}})\citenamefont {Wang}, \citenamefont {Wang},
  \citenamefont {Jiang}, \citenamefont {Zhang}, \citenamefont {Li},
  \citenamefont {Feng}, \citenamefont {Liu}, \citenamefont {Lu}, \citenamefont
  {Sheng}, \citenamefont {Du}, \citenamefont {Gao},\ and\ \citenamefont
  {Xiang}}]{Wang2024}%
  \BibitemOpen
  \bibfield  {author} {\bibinfo {author} {\bibfnamefont {Shasha}\ \bibnamefont
  {Wang}}, \bibinfo {author} {\bibfnamefont {Zhou}\ \bibnamefont {Wang}},
  \bibinfo {author} {\bibfnamefont {Jialiang}\ \bibnamefont {Jiang}}, \bibinfo
  {author} {\bibfnamefont {Ying}\ \bibnamefont {Zhang}}, \bibinfo {author}
  {\bibfnamefont {Ruimin}\ \bibnamefont {Li}}, \bibinfo {author} {\bibfnamefont
  {Yan}\ \bibnamefont {Feng}}, \bibinfo {author} {\bibfnamefont {Ping}\
  \bibnamefont {Liu}}, \bibinfo {author} {\bibfnamefont {Yalin}\ \bibnamefont
  {Lu}}, \bibinfo {author} {\bibfnamefont {Zhigao}\ \bibnamefont {Sheng}},
  \bibinfo {author} {\bibfnamefont {Haifeng}\ \bibnamefont {Du}}, \bibinfo
  {author} {\bibfnamefont {Nan}\ \bibnamefont {Gao}}, \ and\ \bibinfo {author}
  {\bibfnamefont {Bin}\ \bibnamefont {Xiang}},\ }\bibfield  {title} {\enquote
  {\bibinfo {title} {{Temperature-driven spin reorientation transition in van
  der Waals Cr$_{1.7}$Te$_2$ ferromagnet}},}\ }\href {\doibase
  10.1063/5.0202429} {\bibfield  {journal} {\bibinfo  {journal} {Applied
  Physics Letters}\ }\textbf {\bibinfo {volume} {124}} (\bibinfo {year}
  {2024}{\natexlab{a}}),\ 10.1063/5.0202429}\BibitemShut {NoStop}%
\bibitem [{\citenamefont {Jones}\ \emph {et~al.}(2024)\citenamefont {Jones},
  \citenamefont {Das}, \citenamefont {Bhandari}, \citenamefont {Liu},
  \citenamefont {Siegfried}, \citenamefont {Ghimire}, \citenamefont {Tsirkin},
  \citenamefont {Mazin},\ and\ \citenamefont {Ghimire}}]{Jones2024}%
  \BibitemOpen
  \bibfield  {author} {\bibinfo {author} {\bibfnamefont {D.~Connor}\
  \bibnamefont {Jones}}, \bibinfo {author} {\bibfnamefont {Suvadip}\
  \bibnamefont {Das}}, \bibinfo {author} {\bibfnamefont {Hari}\ \bibnamefont
  {Bhandari}}, \bibinfo {author} {\bibfnamefont {Xiaoxiong}\ \bibnamefont
  {Liu}}, \bibinfo {author} {\bibfnamefont {Peter}\ \bibnamefont {Siegfried}},
  \bibinfo {author} {\bibfnamefont {Madhav~P.}\ \bibnamefont {Ghimire}},
  \bibinfo {author} {\bibfnamefont {Stepan~S.}\ \bibnamefont {Tsirkin}},
  \bibinfo {author} {\bibfnamefont {I.~I.}\ \bibnamefont {Mazin}}, \ and\
  \bibinfo {author} {\bibfnamefont {Nirmal~J.}\ \bibnamefont {Ghimire}},\
  }\bibfield  {title} {\enquote {\bibinfo {title} {{Origin of spin
  reorientation and intrinsic anomalous Hall effect in the kagome ferrimagnet
  TbMn$_6$Sn$_6$}},}\ }\href {\doibase 10.1103/physrevb.110.115134} {\bibfield
  {journal} {\bibinfo  {journal} {Physical Review B}\ }\textbf {\bibinfo
  {volume} {110}},\ \bibinfo {pages} {115134} (\bibinfo {year}
  {2024})}\BibitemShut {NoStop}%
\bibitem [{\citenamefont {Lv}\ \emph {et~al.}(2025)\citenamefont {Lv},
  \citenamefont {Zhong}, \citenamefont {Luo}, \citenamefont {Ma}, \citenamefont
  {Chen}, \citenamefont {Wang}, \citenamefont {Luo}, \citenamefont {Gao},
  \citenamefont {Fang},\ and\ \citenamefont {Ren}}]{Lv2025}%
  \BibitemOpen
  \bibfield  {author} {\bibinfo {author} {\bibfnamefont {Bodong}\ \bibnamefont
  {Lv}}, \bibinfo {author} {\bibfnamefont {Rui}\ \bibnamefont {Zhong}},
  \bibinfo {author} {\bibfnamefont {Xiaohua}\ \bibnamefont {Luo}}, \bibinfo
  {author} {\bibfnamefont {Shengcan}\ \bibnamefont {Ma}}, \bibinfo {author}
  {\bibfnamefont {Changcai}\ \bibnamefont {Chen}}, \bibinfo {author}
  {\bibfnamefont {Sujuan}\ \bibnamefont {Wang}}, \bibinfo {author}
  {\bibfnamefont {Qing}\ \bibnamefont {Luo}}, \bibinfo {author} {\bibfnamefont
  {Fei}\ \bibnamefont {Gao}}, \bibinfo {author} {\bibfnamefont {Chunsheng}\
  \bibnamefont {Fang}}, \ and\ \bibinfo {author} {\bibfnamefont {Weijun}\
  \bibnamefont {Ren}},\ }\bibfield  {title} {\enquote {\bibinfo {title}
  {{Anomalous Hall effect and topological Hall effect in Kagome lattice
  material Yb$_{0.90}$Mn$_6$Ge$_{3.25}$Ga$_{0.39}$ single crystal}},}\ }\href
  {\doibase 10.1016/j.scriptamat.2024.116345} {\bibfield  {journal} {\bibinfo
  {journal} {Scripta Materialia}\ }\textbf {\bibinfo {volume} {255}},\ \bibinfo
  {pages} {116345} (\bibinfo {year} {2025})}\BibitemShut {NoStop}%
\bibitem [{\citenamefont {Kumar}\ \emph {et~al.}(2020)\citenamefont {Kumar},
  \citenamefont {Kumar}, \citenamefont {Reehuis}, \citenamefont {Gayles},
  \citenamefont {Sukhanov}, \citenamefont {Hoser}, \citenamefont {Damay},
  \citenamefont {Shekhar}, \citenamefont {Adler},\ and\ \citenamefont
  {Felser}}]{Kumar2020}%
  \BibitemOpen
  \bibfield  {author} {\bibinfo {author} {\bibfnamefont {Vivek}\ \bibnamefont
  {Kumar}}, \bibinfo {author} {\bibfnamefont {Nitesh}\ \bibnamefont {Kumar}},
  \bibinfo {author} {\bibfnamefont {Manfred}\ \bibnamefont {Reehuis}}, \bibinfo
  {author} {\bibfnamefont {Jacob}\ \bibnamefont {Gayles}}, \bibinfo {author}
  {\bibfnamefont {A.~S.}\ \bibnamefont {Sukhanov}}, \bibinfo {author}
  {\bibfnamefont {Andreas}\ \bibnamefont {Hoser}}, \bibinfo {author}
  {\bibfnamefont {Françoise}\ \bibnamefont {Damay}}, \bibinfo {author}
  {\bibfnamefont {Chandra}\ \bibnamefont {Shekhar}}, \bibinfo {author}
  {\bibfnamefont {Peter}\ \bibnamefont {Adler}}, \ and\ \bibinfo {author}
  {\bibfnamefont {Claudia}\ \bibnamefont {Felser}},\ }\bibfield  {title}
  {\enquote {\bibinfo {title} {{Detection of antiskyrmions by topological Hall
  effect in Heusler compounds}},}\ }\href {\doibase
  10.1103/physrevb.101.014424} {\bibfield  {journal} {\bibinfo  {journal}
  {Physical Review B}\ }\textbf {\bibinfo {volume} {101}},\ \bibinfo {pages}
  {014424} (\bibinfo {year} {2020})}\BibitemShut {NoStop}%
\bibitem [{\citenamefont {He}\ \emph {et~al.}(2020)\citenamefont {He},
  \citenamefont {Kroder}, \citenamefont {Gayles}, \citenamefont {Fu},
  \citenamefont {Pan}, \citenamefont {Schnelle}, \citenamefont {Felser},\ and\
  \citenamefont {Fecher}}]{He2020}%
  \BibitemOpen
  \bibfield  {author} {\bibinfo {author} {\bibfnamefont {Yangkun}\ \bibnamefont
  {He}}, \bibinfo {author} {\bibfnamefont {Johannes}\ \bibnamefont {Kroder}},
  \bibinfo {author} {\bibfnamefont {Jacob}\ \bibnamefont {Gayles}}, \bibinfo
  {author} {\bibfnamefont {Chenguang}\ \bibnamefont {Fu}}, \bibinfo {author}
  {\bibfnamefont {Yu}~\bibnamefont {Pan}}, \bibinfo {author} {\bibfnamefont
  {Walter}\ \bibnamefont {Schnelle}}, \bibinfo {author} {\bibfnamefont
  {Claudia}\ \bibnamefont {Felser}}, \ and\ \bibinfo {author} {\bibfnamefont
  {Gerhard~H.}\ \bibnamefont {Fecher}},\ }\bibfield  {title} {\enquote
  {\bibinfo {title} {{Large topological Hall effect in an easy-cone ferromagnet
  (Cr$_{0.9}$B$_{0.1}$)Te}},}\ }\href {\doibase 10.1063/5.0018229} {\bibfield
  {journal} {\bibinfo  {journal} {Applied Physics Letters}\ }\textbf {\bibinfo
  {volume} {117}} (\bibinfo {year} {2020}),\ 10.1063/5.0018229}\BibitemShut
  {NoStop}%
\bibitem [{\citenamefont {Malaman}\ \emph {et~al.}(1976)\citenamefont
  {Malaman}, \citenamefont {Roques}, \citenamefont {Courtois},\ and\
  \citenamefont {Protas}}]{Malaman1976}%
  \BibitemOpen
  \bibfield  {author} {\bibinfo {author} {\bibfnamefont {B.}~\bibnamefont
  {Malaman}}, \bibinfo {author} {\bibfnamefont {B.}~\bibnamefont {Roques}},
  \bibinfo {author} {\bibfnamefont {A.}~\bibnamefont {Courtois}}, \ and\
  \bibinfo {author} {\bibfnamefont {J.}~\bibnamefont {Protas}},\ }\bibfield
  {title} {\enquote {\bibinfo {title} {{Structure cristalline du stannure de
  fer Fe$_3$Sn$_2$}},}\ }\href {\doibase 10.1107/s0567740876005323} {\bibfield
  {journal} {\bibinfo  {journal} {Acta Crystallographica Section B Structural
  Crystallography and Crystal Chemistry}\ }\textbf {\bibinfo {volume} {32}},\
  \bibinfo {pages} {1348--1351} (\bibinfo {year} {1976})}\BibitemShut {NoStop}%
\bibitem [{\citenamefont {Malaman}\ \emph {et~al.}(1978)\citenamefont
  {Malaman}, \citenamefont {Fruchart},\ and\ \citenamefont
  {Caer}}]{Malaman1978}%
  \BibitemOpen
  \bibfield  {author} {\bibinfo {author} {\bibfnamefont {B}~\bibnamefont
  {Malaman}}, \bibinfo {author} {\bibfnamefont {D}~\bibnamefont {Fruchart}}, \
  and\ \bibinfo {author} {\bibfnamefont {G~Le}\ \bibnamefont {Caer}},\
  }\bibfield  {title} {\enquote {\bibinfo {title} {{Magnetic properties of
  Fe$_3$Sn$_2$. II. Neutron diffraction study (and Mossbauer effect)}},}\
  }\href {\doibase 10.1088/0305-4608/8/11/022} {\bibfield  {journal} {\bibinfo
  {journal} {Journal of Physics F: Metal Physics}\ }\textbf {\bibinfo {volume}
  {8}},\ \bibinfo {pages} {2389--2399} (\bibinfo {year} {1978})}\BibitemShut
  {NoStop}%
\bibitem [{\citenamefont {He}\ \emph {et~al.}(2021)\citenamefont {He},
  \citenamefont {Peis}, \citenamefont {Stumberger}, \citenamefont {Prodan},
  \citenamefont {Tsurkan}, \citenamefont {Unglert}, \citenamefont {Chioncel},
  \citenamefont {Kézsmárki},\ and\ \citenamefont {Hackl}}]{He2021}%
  \BibitemOpen
  \bibfield  {author} {\bibinfo {author} {\bibfnamefont {Ge}~\bibnamefont
  {He}}, \bibinfo {author} {\bibfnamefont {Leander}\ \bibnamefont {Peis}},
  \bibinfo {author} {\bibfnamefont {Ramona}\ \bibnamefont {Stumberger}},
  \bibinfo {author} {\bibfnamefont {Lilian}\ \bibnamefont {Prodan}}, \bibinfo
  {author} {\bibfnamefont {Vladimir}\ \bibnamefont {Tsurkan}}, \bibinfo
  {author} {\bibfnamefont {Nico}\ \bibnamefont {Unglert}}, \bibinfo {author}
  {\bibfnamefont {Liviu}\ \bibnamefont {Chioncel}}, \bibinfo {author}
  {\bibfnamefont {István}\ \bibnamefont {Kézsmárki}}, \ and\ \bibinfo
  {author} {\bibfnamefont {Rudi}\ \bibnamefont {Hackl}},\ }\bibfield  {title}
  {\enquote {\bibinfo {title} {{Phonon Anomalies Associated with Spin
  Reorientation in the Kagome Ferromagnet Fe$_3$Sn$_2$}},}\ }\href {\doibase
  10.1002/pssb.202100169} {\bibfield  {journal} {\bibinfo  {journal} {physica
  status solidi (b)}\ }\textbf {\bibinfo {volume} {259}} (\bibinfo {year}
  {2021}),\ 10.1002/pssb.202100169}\BibitemShut {NoStop}%
\bibitem [{\citenamefont {Ye}\ \emph {et~al.}(2018)\citenamefont {Ye},
  \citenamefont {Kang}, \citenamefont {Liu}, \citenamefont {von Cube},
  \citenamefont {Wicker}, \citenamefont {Suzuki}, \citenamefont {Jozwiak},
  \citenamefont {Bostwick}, \citenamefont {Rotenberg}, \citenamefont {Bell},
  \citenamefont {Fu}, \citenamefont {Comin},\ and\ \citenamefont
  {Checkelsky}}]{Ye2018}%
  \BibitemOpen
  \bibfield  {author} {\bibinfo {author} {\bibfnamefont {Linda}\ \bibnamefont
  {Ye}}, \bibinfo {author} {\bibfnamefont {Mingu}\ \bibnamefont {Kang}},
  \bibinfo {author} {\bibfnamefont {Junwei}\ \bibnamefont {Liu}}, \bibinfo
  {author} {\bibfnamefont {Felix}\ \bibnamefont {von Cube}}, \bibinfo {author}
  {\bibfnamefont {Christina~R.}\ \bibnamefont {Wicker}}, \bibinfo {author}
  {\bibfnamefont {Takehito}\ \bibnamefont {Suzuki}}, \bibinfo {author}
  {\bibfnamefont {Chris}\ \bibnamefont {Jozwiak}}, \bibinfo {author}
  {\bibfnamefont {Aaron}\ \bibnamefont {Bostwick}}, \bibinfo {author}
  {\bibfnamefont {Eli}\ \bibnamefont {Rotenberg}}, \bibinfo {author}
  {\bibfnamefont {David~C.}\ \bibnamefont {Bell}}, \bibinfo {author}
  {\bibfnamefont {Liang}\ \bibnamefont {Fu}}, \bibinfo {author} {\bibfnamefont
  {Riccardo}\ \bibnamefont {Comin}}, \ and\ \bibinfo {author} {\bibfnamefont
  {Joseph~G.}\ \bibnamefont {Checkelsky}},\ }\bibfield  {title} {\enquote
  {\bibinfo {title} {{Massive Dirac fermions in a ferromagnetic kagome
  metal}},}\ }\href {\doibase 10.1038/nature25987} {\bibfield  {journal}
  {\bibinfo  {journal} {Nature}\ }\textbf {\bibinfo {volume} {555}},\ \bibinfo
  {pages} {638--642} (\bibinfo {year} {2018})}\BibitemShut {NoStop}%
\bibitem [{\citenamefont {Biswas}\ \emph {et~al.}(2020)\citenamefont {Biswas},
  \citenamefont {Iakutkina}, \citenamefont {Wang}, \citenamefont {Lei},
  \citenamefont {Dressel},\ and\ \citenamefont {Uykur}}]{Biswas2020}%
  \BibitemOpen
  \bibfield  {author} {\bibinfo {author} {\bibfnamefont {A.}~\bibnamefont
  {Biswas}}, \bibinfo {author} {\bibfnamefont {O.}~\bibnamefont {Iakutkina}},
  \bibinfo {author} {\bibfnamefont {Q.}~\bibnamefont {Wang}}, \bibinfo {author}
  {\bibfnamefont {H.C.}\ \bibnamefont {Lei}}, \bibinfo {author} {\bibfnamefont
  {M.}~\bibnamefont {Dressel}}, \ and\ \bibinfo {author} {\bibfnamefont
  {E.}~\bibnamefont {Uykur}},\ }\bibfield  {title} {\enquote {\bibinfo {title}
  {{Spin-Reorientation-Induced Band Gap in Fe$_3$Sn$_2$: Optical Signatures of
  Weyl Nodes}},}\ }\href {\doibase 10.1103/physrevlett.125.076403} {\bibfield
  {journal} {\bibinfo  {journal} {Physical Review Letters}\ }\textbf {\bibinfo
  {volume} {125}},\ \bibinfo {pages} {076403} (\bibinfo {year}
  {2020})}\BibitemShut {NoStop}%
\bibitem [{\citenamefont {Lin}\ \emph {et~al.}(2018{\natexlab{b}})\citenamefont
  {Lin}, \citenamefont {Choi}, \citenamefont {Zhang}, \citenamefont {Qin},
  \citenamefont {Yi}, \citenamefont {Wang}, \citenamefont {Li}, \citenamefont
  {Wang}, \citenamefont {Zhang}, \citenamefont {Sun}, \citenamefont {Wei},
  \citenamefont {Zhang}, \citenamefont {Guo}, \citenamefont {Lu}, \citenamefont
  {Cho}, \citenamefont {Zeng},\ and\ \citenamefont {Zhang}}]{Lin2018a}%
  \BibitemOpen
  \bibfield  {author} {\bibinfo {author} {\bibfnamefont {Zhiyong}\ \bibnamefont
  {Lin}}, \bibinfo {author} {\bibfnamefont {Jin-Ho}\ \bibnamefont {Choi}},
  \bibinfo {author} {\bibfnamefont {Qiang}\ \bibnamefont {Zhang}}, \bibinfo
  {author} {\bibfnamefont {Wei}\ \bibnamefont {Qin}}, \bibinfo {author}
  {\bibfnamefont {Seho}\ \bibnamefont {Yi}}, \bibinfo {author} {\bibfnamefont
  {Pengdong}\ \bibnamefont {Wang}}, \bibinfo {author} {\bibfnamefont {Lin}\
  \bibnamefont {Li}}, \bibinfo {author} {\bibfnamefont {Yifan}\ \bibnamefont
  {Wang}}, \bibinfo {author} {\bibfnamefont {Hui}\ \bibnamefont {Zhang}},
  \bibinfo {author} {\bibfnamefont {Zhe}\ \bibnamefont {Sun}}, \bibinfo
  {author} {\bibfnamefont {Laiming}\ \bibnamefont {Wei}}, \bibinfo {author}
  {\bibfnamefont {Shengbai}\ \bibnamefont {Zhang}}, \bibinfo {author}
  {\bibfnamefont {Tengfei}\ \bibnamefont {Guo}}, \bibinfo {author}
  {\bibfnamefont {Qingyou}\ \bibnamefont {Lu}}, \bibinfo {author}
  {\bibfnamefont {Jun-Hyung}\ \bibnamefont {Cho}}, \bibinfo {author}
  {\bibfnamefont {Changgan}\ \bibnamefont {Zeng}}, \ and\ \bibinfo {author}
  {\bibfnamefont {Zhenyu}\ \bibnamefont {Zhang}},\ }\bibfield  {title}
  {\enquote {\bibinfo {title} {{Flatbands and Emergent Ferromagnetic Ordering
  in Fe$_3$Sn$_2$ Kagome Lattices}},}\ }\href {\doibase
  10.1103/physrevlett.121.096401} {\bibfield  {journal} {\bibinfo  {journal}
  {Physical Review Letters}\ }\textbf {\bibinfo {volume} {121}},\ \bibinfo
  {pages} {096401} (\bibinfo {year} {2018}{\natexlab{b}})}\BibitemShut
  {NoStop}%
\bibitem [{\citenamefont {Schilberth}\ \emph {et~al.}(2022)\citenamefont
  {Schilberth}, \citenamefont {Unglert}, \citenamefont {Prodan}, \citenamefont
  {Meggle}, \citenamefont {Ebad~Allah}, \citenamefont {Kuntscher},
  \citenamefont {Tsirlin}, \citenamefont {Tsurkan}, \citenamefont
  {Deisenhofer}, \citenamefont {Chioncel}, \citenamefont {Kézsmárki},\ and\
  \citenamefont {Bordács}}]{Schilberth2022}%
  \BibitemOpen
  \bibfield  {author} {\bibinfo {author} {\bibfnamefont {F.}~\bibnamefont
  {Schilberth}}, \bibinfo {author} {\bibfnamefont {N.}~\bibnamefont {Unglert}},
  \bibinfo {author} {\bibfnamefont {L.}~\bibnamefont {Prodan}}, \bibinfo
  {author} {\bibfnamefont {F.}~\bibnamefont {Meggle}}, \bibinfo {author}
  {\bibfnamefont {J.}~\bibnamefont {Ebad~Allah}}, \bibinfo {author}
  {\bibfnamefont {C.~A.}\ \bibnamefont {Kuntscher}}, \bibinfo {author}
  {\bibfnamefont {A.~A.}\ \bibnamefont {Tsirlin}}, \bibinfo {author}
  {\bibfnamefont {V.}~\bibnamefont {Tsurkan}}, \bibinfo {author} {\bibfnamefont
  {J.}~\bibnamefont {Deisenhofer}}, \bibinfo {author} {\bibfnamefont
  {L.}~\bibnamefont {Chioncel}}, \bibinfo {author} {\bibfnamefont
  {I.}~\bibnamefont {Kézsmárki}}, \ and\ \bibinfo {author} {\bibfnamefont
  {S.}~\bibnamefont {Bordács}},\ }\bibfield  {title} {\enquote {\bibinfo
  {title} {{Magneto-optical detection of topological contributions to the
  anomalous Hall effect in a kagome ferromagnet}},}\ }\href {\doibase
  10.1103/physrevb.106.144404} {\bibfield  {journal} {\bibinfo  {journal}
  {Phys. Rev. B}\ }\textbf {\bibinfo {volume} {106}},\ \bibinfo {pages}
  {144404} (\bibinfo {year} {2022})}\BibitemShut {NoStop}%
\bibitem [{\citenamefont {Du}\ \emph {et~al.}(2022)\citenamefont {Du},
  \citenamefont {Hu}, \citenamefont {Han}, \citenamefont {Camino},
  \citenamefont {Zhu},\ and\ \citenamefont {Petrovic}}]{Du2022}%
  \BibitemOpen
  \bibfield  {author} {\bibinfo {author} {\bibfnamefont {Qianheng}\
  \bibnamefont {Du}}, \bibinfo {author} {\bibfnamefont {Zhixiang}\ \bibnamefont
  {Hu}}, \bibinfo {author} {\bibfnamefont {Myung-Geun}\ \bibnamefont {Han}},
  \bibinfo {author} {\bibfnamefont {Fernando}\ \bibnamefont {Camino}}, \bibinfo
  {author} {\bibfnamefont {Yimei}\ \bibnamefont {Zhu}}, \ and\ \bibinfo
  {author} {\bibfnamefont {C.}~\bibnamefont {Petrovic}},\ }\bibfield  {title}
  {\enquote {\bibinfo {title} {{Topological Hall Effect Anisotropy in Kagome
  Bilayer Metal Fe$_3$Sn$_2$}},}\ }\href {\doibase
  10.1103/physrevlett.129.236601} {\bibfield  {journal} {\bibinfo  {journal}
  {Phys. Rev. Lett.}\ }\textbf {\bibinfo {volume} {129}},\ \bibinfo {pages}
  {236601} (\bibinfo {year} {2022})}\BibitemShut {NoStop}%
\bibitem [{\citenamefont {Hou}\ \emph {et~al.}(2018)\citenamefont {Hou},
  \citenamefont {Ren}, \citenamefont {Ding}, \citenamefont {Xu}, \citenamefont
  {Wang}, \citenamefont {Yang}, \citenamefont {Zhang}, \citenamefont {Zhang},
  \citenamefont {Liu}, \citenamefont {Xu}, \citenamefont {Wang}, \citenamefont
  {Wu}, \citenamefont {Zhang}, \citenamefont {Shen},\ and\ \citenamefont
  {Zhang}}]{Hou2018}%
  \BibitemOpen
  \bibfield  {author} {\bibinfo {author} {\bibfnamefont {Zhipeng}\ \bibnamefont
  {Hou}}, \bibinfo {author} {\bibfnamefont {Weijun}\ \bibnamefont {Ren}},
  \bibinfo {author} {\bibfnamefont {Bei}\ \bibnamefont {Ding}}, \bibinfo
  {author} {\bibfnamefont {Guizhou}\ \bibnamefont {Xu}}, \bibinfo {author}
  {\bibfnamefont {Yue}\ \bibnamefont {Wang}}, \bibinfo {author} {\bibfnamefont
  {Bing}\ \bibnamefont {Yang}}, \bibinfo {author} {\bibfnamefont {Qiang}\
  \bibnamefont {Zhang}}, \bibinfo {author} {\bibfnamefont {Ying}\ \bibnamefont
  {Zhang}}, \bibinfo {author} {\bibfnamefont {Enke}\ \bibnamefont {Liu}},
  \bibinfo {author} {\bibfnamefont {Feng}\ \bibnamefont {Xu}}, \bibinfo
  {author} {\bibfnamefont {Wenhong}\ \bibnamefont {Wang}}, \bibinfo {author}
  {\bibfnamefont {Guangheng}\ \bibnamefont {Wu}}, \bibinfo {author}
  {\bibfnamefont {Xixiang}\ \bibnamefont {Zhang}}, \bibinfo {author}
  {\bibfnamefont {Baogen}\ \bibnamefont {Shen}}, \ and\ \bibinfo {author}
  {\bibfnamefont {Zhidong}\ \bibnamefont {Zhang}},\ }\bibfield  {title}
  {\enquote {\bibinfo {title} {Observation of various and spontaneous magnetic
  skyrmionic bubbles at room temperature in a frustrated kagome magnet with
  uniaxial magnetic anisotropy},}\ }\href {\doibase 10.1002/adma.201706306}
  {\bibfield  {journal} {\bibinfo  {journal} {Adv. Mater.}\ }\textbf {\bibinfo
  {volume} {30}} (\bibinfo {year} {2018}),\ 10.1002/adma.201706306}\BibitemShut
  {NoStop}%
\bibitem [{\citenamefont {Altthaler}\ \emph {et~al.}(2021)\citenamefont
  {Altthaler}, \citenamefont {Lysne}, \citenamefont {Roede}, \citenamefont
  {Prodan}, \citenamefont {Tsurkan}, \citenamefont {Kassem}, \citenamefont
  {Nakamura}, \citenamefont {Krohns}, \citenamefont {Kézsmárki},\ and\
  \citenamefont {Meier}}]{Altthaler2021}%
  \BibitemOpen
  \bibfield  {author} {\bibinfo {author} {\bibfnamefont {Markus}\ \bibnamefont
  {Altthaler}}, \bibinfo {author} {\bibfnamefont {Erik}\ \bibnamefont {Lysne}},
  \bibinfo {author} {\bibfnamefont {Erik}\ \bibnamefont {Roede}}, \bibinfo
  {author} {\bibfnamefont {Lilian}\ \bibnamefont {Prodan}}, \bibinfo {author}
  {\bibfnamefont {Vladimir}\ \bibnamefont {Tsurkan}}, \bibinfo {author}
  {\bibfnamefont {Mohamed~A.}\ \bibnamefont {Kassem}}, \bibinfo {author}
  {\bibfnamefont {Hiroyuki}\ \bibnamefont {Nakamura}}, \bibinfo {author}
  {\bibfnamefont {Stephan}\ \bibnamefont {Krohns}}, \bibinfo {author}
  {\bibfnamefont {István}\ \bibnamefont {Kézsmárki}}, \ and\ \bibinfo
  {author} {\bibfnamefont {Dennis}\ \bibnamefont {Meier}},\ }\bibfield  {title}
  {\enquote {\bibinfo {title} {{Magnetic and geometric control of spin textures
  in the itinerant kagome magnet Fe$_3$Sn$_2$}},}\ }\href {\doibase
  10.1103/physrevresearch.3.043191} {\bibfield  {journal} {\bibinfo  {journal}
  {Physical Review Research}\ }\textbf {\bibinfo {volume} {3}},\ \bibinfo
  {pages} {043191} (\bibinfo {year} {2021})}\BibitemShut {NoStop}%
\bibitem [{\citenamefont {Ekahana}\ \emph {et~al.}(2024)\citenamefont
  {Ekahana}, \citenamefont {Soh}, \citenamefont {Tamai}, \citenamefont
  {Gosálbez-Martínez}, \citenamefont {Yao}, \citenamefont {Hunter},
  \citenamefont {Fan}, \citenamefont {Wang}, \citenamefont {Li}, \citenamefont
  {Kleibert}, \citenamefont {Vaz}, \citenamefont {Ma}, \citenamefont {Lee},
  \citenamefont {Xiong}, \citenamefont {Yazyev}, \citenamefont {Baumberger},
  \citenamefont {Shi},\ and\ \citenamefont {Aeppli}}]{Ekahana2024}%
  \BibitemOpen
  \bibfield  {author} {\bibinfo {author} {\bibfnamefont {Sandy~Adhitia}\
  \bibnamefont {Ekahana}}, \bibinfo {author} {\bibfnamefont {Y.}~\bibnamefont
  {Soh}}, \bibinfo {author} {\bibfnamefont {Anna}\ \bibnamefont {Tamai}},
  \bibinfo {author} {\bibfnamefont {Daniel}\ \bibnamefont
  {Gosálbez-Martínez}}, \bibinfo {author} {\bibfnamefont {Mengyu}\
  \bibnamefont {Yao}}, \bibinfo {author} {\bibfnamefont {Andrew}\ \bibnamefont
  {Hunter}}, \bibinfo {author} {\bibfnamefont {Wenhui}\ \bibnamefont {Fan}},
  \bibinfo {author} {\bibfnamefont {Yihao}\ \bibnamefont {Wang}}, \bibinfo
  {author} {\bibfnamefont {Junbo}\ \bibnamefont {Li}}, \bibinfo {author}
  {\bibfnamefont {Armin}\ \bibnamefont {Kleibert}}, \bibinfo {author}
  {\bibfnamefont {C.~A.~F.}\ \bibnamefont {Vaz}}, \bibinfo {author}
  {\bibfnamefont {Junzhang}\ \bibnamefont {Ma}}, \bibinfo {author}
  {\bibfnamefont {Hyungjun}\ \bibnamefont {Lee}}, \bibinfo {author}
  {\bibfnamefont {Yimin}\ \bibnamefont {Xiong}}, \bibinfo {author}
  {\bibfnamefont {Oleg~V.}\ \bibnamefont {Yazyev}}, \bibinfo {author}
  {\bibfnamefont {Felix}\ \bibnamefont {Baumberger}}, \bibinfo {author}
  {\bibfnamefont {Ming}\ \bibnamefont {Shi}}, \ and\ \bibinfo {author}
  {\bibfnamefont {G.}~\bibnamefont {Aeppli}},\ }\bibfield  {title} {\enquote
  {\bibinfo {title} {{Anomalous electrons in a metallic kagome ferromagnet}},}\
  }\href {\doibase 10.1038/s41586-024-07085-w} {\bibfield  {journal} {\bibinfo
  {journal} {Nature}\ }\textbf {\bibinfo {volume} {627}},\ \bibinfo {pages}
  {67--72} (\bibinfo {year} {2024})}\BibitemShut {NoStop}%
\bibitem [{\citenamefont {Gonçalves-Faria}\ \emph {et~al.}(2024)\citenamefont
  {Gonçalves-Faria}, \citenamefont {Pashkin}, \citenamefont {Wang},
  \citenamefont {Lei}, \citenamefont {Winnerl}, \citenamefont {Tsirlin},
  \citenamefont {Helm},\ and\ \citenamefont {Uykur}}]{GoncalvesFaria2024}%
  \BibitemOpen
  \bibfield  {author} {\bibinfo {author} {\bibfnamefont {Marcos~V.}\
  \bibnamefont {Gonçalves-Faria}}, \bibinfo {author} {\bibfnamefont {Alexej}\
  \bibnamefont {Pashkin}}, \bibinfo {author} {\bibfnamefont {Qi}~\bibnamefont
  {Wang}}, \bibinfo {author} {\bibfnamefont {Hechang~C.}\ \bibnamefont {Lei}},
  \bibinfo {author} {\bibfnamefont {Stephan}\ \bibnamefont {Winnerl}}, \bibinfo
  {author} {\bibfnamefont {Alexander~A.}\ \bibnamefont {Tsirlin}}, \bibinfo
  {author} {\bibfnamefont {Manfred}\ \bibnamefont {Helm}}, \ and\ \bibinfo
  {author} {\bibfnamefont {Ece}\ \bibnamefont {Uykur}},\ }\bibfield  {title}
  {\enquote {\bibinfo {title} {Coherent phonon and unconventional carriers in
  the magnetic kagome metal fe$_3$sn$_2$},}\ }\href {\doibase
  10.1038/s41535-024-00642-6} {\bibfield  {journal} {\bibinfo  {journal} {npj
  Quantum Materials}\ }\textbf {\bibinfo {volume} {9}} (\bibinfo {year}
  {2024}),\ 10.1038/s41535-024-00642-6}\BibitemShut {NoStop}%
\bibitem [{\citenamefont {Zhang}\ \emph {et~al.}(2024)\citenamefont {Zhang},
  \citenamefont {Asmara}, \citenamefont {Tseng}, \citenamefont {Li},
  \citenamefont {Xiong}, \citenamefont {Wei}, \citenamefont {Yu}, \citenamefont
  {Galdino}, \citenamefont {Zhang}, \citenamefont {Kummer}, \citenamefont
  {Strocov}, \citenamefont {Soh}, \citenamefont {Schmitt},\ and\ \citenamefont
  {Aeppli}}]{Zhang2024}%
  \BibitemOpen
  \bibfield  {author} {\bibinfo {author} {\bibfnamefont {Wenliang}\
  \bibnamefont {Zhang}}, \bibinfo {author} {\bibfnamefont {Teguh~Citra}\
  \bibnamefont {Asmara}}, \bibinfo {author} {\bibfnamefont {Yi}~\bibnamefont
  {Tseng}}, \bibinfo {author} {\bibfnamefont {Junbo}\ \bibnamefont {Li}},
  \bibinfo {author} {\bibfnamefont {Yimin}\ \bibnamefont {Xiong}}, \bibinfo
  {author} {\bibfnamefont {Yuan}\ \bibnamefont {Wei}}, \bibinfo {author}
  {\bibfnamefont {Tianlun}\ \bibnamefont {Yu}}, \bibinfo {author}
  {\bibfnamefont {Carlos~William}\ \bibnamefont {Galdino}}, \bibinfo {author}
  {\bibfnamefont {Zhijia}\ \bibnamefont {Zhang}}, \bibinfo {author}
  {\bibfnamefont {Kurt}\ \bibnamefont {Kummer}}, \bibinfo {author}
  {\bibfnamefont {Vladimir~N.}\ \bibnamefont {Strocov}}, \bibinfo {author}
  {\bibfnamefont {Y.}~\bibnamefont {Soh}}, \bibinfo {author} {\bibfnamefont
  {Thorsten}\ \bibnamefont {Schmitt}}, \ and\ \bibinfo {author} {\bibfnamefont
  {Gabriel}\ \bibnamefont {Aeppli}},\ }\bibfield  {title} {\enquote {\bibinfo
  {title} {{Spin waves and orbital contribution to ferromagnetism in a
  topological metal}},}\ }\href {\doibase 10.1038/s41467-024-53152-1}
  {\bibfield  {journal} {\bibinfo  {journal} {Nature Communications}\ }\textbf
  {\bibinfo {volume} {15}} (\bibinfo {year} {2024}),\
  10.1038/s41467-024-53152-1}\BibitemShut {NoStop}%
\bibitem [{\citenamefont {Wang}\ \emph
  {et~al.}(2024{\natexlab{b}})\citenamefont {Wang}, \citenamefont {Zhu},
  \citenamefont {Chen}, \citenamefont {Wang}, \citenamefont {Liu},
  \citenamefont {Huang}, \citenamefont {Jiang}, \citenamefont {Zhao},
  \citenamefont {Shi}, \citenamefont {Tian}, \citenamefont {Wang},
  \citenamefont {Yao}, \citenamefont {Yu}, \citenamefont {Wang}, \citenamefont
  {Xiao}, \citenamefont {Yang},\ and\ \citenamefont {Wu}}]{Wang2024_2}%
  \BibitemOpen
  \bibfield  {author} {\bibinfo {author} {\bibfnamefont {Lujunyu}\ \bibnamefont
  {Wang}}, \bibinfo {author} {\bibfnamefont {Jiaojiao}\ \bibnamefont {Zhu}},
  \bibinfo {author} {\bibfnamefont {Haiyun}\ \bibnamefont {Chen}}, \bibinfo
  {author} {\bibfnamefont {Hui}\ \bibnamefont {Wang}}, \bibinfo {author}
  {\bibfnamefont {Jinjin}\ \bibnamefont {Liu}}, \bibinfo {author}
  {\bibfnamefont {Yue-Xin}\ \bibnamefont {Huang}}, \bibinfo {author}
  {\bibfnamefont {Bingyan}\ \bibnamefont {Jiang}}, \bibinfo {author}
  {\bibfnamefont {Jiaji}\ \bibnamefont {Zhao}}, \bibinfo {author}
  {\bibfnamefont {Hengjie}\ \bibnamefont {Shi}}, \bibinfo {author}
  {\bibfnamefont {Guang}\ \bibnamefont {Tian}}, \bibinfo {author}
  {\bibfnamefont {Haoyu}\ \bibnamefont {Wang}}, \bibinfo {author}
  {\bibfnamefont {Yugui}\ \bibnamefont {Yao}}, \bibinfo {author} {\bibfnamefont
  {Dapeng}\ \bibnamefont {Yu}}, \bibinfo {author} {\bibfnamefont {Zhiwei}\
  \bibnamefont {Wang}}, \bibinfo {author} {\bibfnamefont {Cong}\ \bibnamefont
  {Xiao}}, \bibinfo {author} {\bibfnamefont {Shengyuan~A.}\ \bibnamefont
  {Yang}}, \ and\ \bibinfo {author} {\bibfnamefont {Xiaosong}\ \bibnamefont
  {Wu}},\ }\bibfield  {title} {\enquote {\bibinfo {title} {{Orbital
  Magneto-Nonlinear Anomalous Hall Effect in Kagome Magnet Fe$_3$Sn$_2$}},}\
  }\href {\doibase 10.1103/physrevlett.132.106601} {\bibfield  {journal}
  {\bibinfo  {journal} {Physical Review Letters}\ }\textbf {\bibinfo {volume}
  {132}},\ \bibinfo {pages} {106601} (\bibinfo {year}
  {2024}{\natexlab{b}})}\BibitemShut {NoStop}%
\bibitem [{\citenamefont {Fenner}\ \emph {et~al.}(2009)\citenamefont {Fenner},
  \citenamefont {Dee},\ and\ \citenamefont {Wills}}]{Fenner2009}%
  \BibitemOpen
  \bibfield  {author} {\bibinfo {author} {\bibfnamefont {L~A}\ \bibnamefont
  {Fenner}}, \bibinfo {author} {\bibfnamefont {A~A}\ \bibnamefont {Dee}}, \
  and\ \bibinfo {author} {\bibfnamefont {A~S}\ \bibnamefont {Wills}},\
  }\bibfield  {title} {\enquote {\bibinfo {title} {{Non-collinearity and spin
  frustration in the itinerant kagome ferromagnet Fe$_3$Sn$_2$}},}\ }\href
  {\doibase 10.1088/0953-8984/21/45/452202} {\bibfield  {journal} {\bibinfo
  {journal} {Journal of Physics: Condensed Matter}\ }\textbf {\bibinfo {volume}
  {21}},\ \bibinfo {pages} {452202} (\bibinfo {year} {2009})}\BibitemShut
  {NoStop}%
\bibitem [{\citenamefont {Fayyazi}\ \emph {et~al.}(2019)\citenamefont
  {Fayyazi}, \citenamefont {Skokov}, \citenamefont {Faske}, \citenamefont
  {Opahle}, \citenamefont {Duerrschnabel}, \citenamefont {Helbig},
  \citenamefont {Soldatov}, \citenamefont {Rohrmann}, \citenamefont
  {Molina-Luna}, \citenamefont {Güth}, \citenamefont {Zhang}, \citenamefont
  {Donner}, \citenamefont {Schäfer},\ and\ \citenamefont
  {Gutfleisch}}]{Fayyazi2019}%
  \BibitemOpen
  \bibfield  {author} {\bibinfo {author} {\bibfnamefont {Bahar}\ \bibnamefont
  {Fayyazi}}, \bibinfo {author} {\bibfnamefont {Konstantin~P.}\ \bibnamefont
  {Skokov}}, \bibinfo {author} {\bibfnamefont {Tom}\ \bibnamefont {Faske}},
  \bibinfo {author} {\bibfnamefont {Ingo}\ \bibnamefont {Opahle}}, \bibinfo
  {author} {\bibfnamefont {Michael}\ \bibnamefont {Duerrschnabel}}, \bibinfo
  {author} {\bibfnamefont {Tim}\ \bibnamefont {Helbig}}, \bibinfo {author}
  {\bibfnamefont {Ivan}\ \bibnamefont {Soldatov}}, \bibinfo {author}
  {\bibfnamefont {Urban}\ \bibnamefont {Rohrmann}}, \bibinfo {author}
  {\bibfnamefont {Leopoldo}\ \bibnamefont {Molina-Luna}}, \bibinfo {author}
  {\bibfnamefont {Konrad}\ \bibnamefont {Güth}}, \bibinfo {author}
  {\bibfnamefont {Hongbin}\ \bibnamefont {Zhang}}, \bibinfo {author}
  {\bibfnamefont {Wolfgang}\ \bibnamefont {Donner}}, \bibinfo {author}
  {\bibfnamefont {Rudolf}\ \bibnamefont {Schäfer}}, \ and\ \bibinfo {author}
  {\bibfnamefont {Oliver}\ \bibnamefont {Gutfleisch}},\ }\bibfield  {title}
  {\enquote {\bibinfo {title} {{Experimental and computational analysis of
  binary Fe-Sn ferromagnetic compounds}},}\ }\href {\doibase
  10.1016/j.actamat.2019.08.054} {\bibfield  {journal} {\bibinfo  {journal}
  {Acta Materialia}\ }\textbf {\bibinfo {volume} {180}},\ \bibinfo {pages}
  {126--140} (\bibinfo {year} {2019})}\BibitemShut {NoStop}%
\bibitem [{\citenamefont {Heritage}\ \emph {et~al.}(2020)\citenamefont
  {Heritage}, \citenamefont {Bryant}, \citenamefont {Fenner}, \citenamefont
  {Wills}, \citenamefont {Aeppli},\ and\ \citenamefont {Soh}}]{Heritage2020}%
  \BibitemOpen
  \bibfield  {author} {\bibinfo {author} {\bibfnamefont {Kevin}\ \bibnamefont
  {Heritage}}, \bibinfo {author} {\bibfnamefont {Ben}\ \bibnamefont {Bryant}},
  \bibinfo {author} {\bibfnamefont {Laura~A.}\ \bibnamefont {Fenner}}, \bibinfo
  {author} {\bibfnamefont {Andrew~S.}\ \bibnamefont {Wills}}, \bibinfo {author}
  {\bibfnamefont {Gabriel}\ \bibnamefont {Aeppli}}, \ and\ \bibinfo {author}
  {\bibfnamefont {Yeong‐Ah}\ \bibnamefont {Soh}},\ }\bibfield  {title}
  {\enquote {\bibinfo {title} {{Images of a First‐Order Spin‐Reorientation
  Phase Transition in a Metallic Kagome Ferromagnet}},}\ }\href {\doibase
  10.1002/adfm.201909163} {\bibfield  {journal} {\bibinfo  {journal} {Advanced
  Functional Materials}\ }\textbf {\bibinfo {volume} {30}} (\bibinfo {year}
  {2020}),\ 10.1002/adfm.201909163}\BibitemShut {NoStop}%
\bibitem [{\citenamefont {Wu}\ \emph {et~al.}(2021)\citenamefont {Wu},
  \citenamefont {Song}, \citenamefont {Yu}, \citenamefont {Wang}, \citenamefont
  {Xia}, \citenamefont {Hong}, \citenamefont {Zu}, \citenamefont {Du},
  \citenamefont {Vallobra}, \citenamefont {Liu}, \citenamefont {Torii},
  \citenamefont {Kamiyama}, \citenamefont {Xiong},\ and\ \citenamefont
  {Zhao}}]{Wu2021}%
  \BibitemOpen
  \bibfield  {author} {\bibinfo {author} {\bibfnamefont {Peng}\ \bibnamefont
  {Wu}}, \bibinfo {author} {\bibfnamefont {Jiuhui}\ \bibnamefont {Song}},
  \bibinfo {author} {\bibfnamefont {Xiaoxiang}\ \bibnamefont {Yu}}, \bibinfo
  {author} {\bibfnamefont {Yihao}\ \bibnamefont {Wang}}, \bibinfo {author}
  {\bibfnamefont {Kang}\ \bibnamefont {Xia}}, \bibinfo {author} {\bibfnamefont
  {Bin}\ \bibnamefont {Hong}}, \bibinfo {author} {\bibfnamefont {Lin}\
  \bibnamefont {Zu}}, \bibinfo {author} {\bibfnamefont {Yinchang}\ \bibnamefont
  {Du}}, \bibinfo {author} {\bibfnamefont {Pierre}\ \bibnamefont {Vallobra}},
  \bibinfo {author} {\bibfnamefont {Fengguang}\ \bibnamefont {Liu}}, \bibinfo
  {author} {\bibfnamefont {Shuki}\ \bibnamefont {Torii}}, \bibinfo {author}
  {\bibfnamefont {Takashi}\ \bibnamefont {Kamiyama}}, \bibinfo {author}
  {\bibfnamefont {Yimin}\ \bibnamefont {Xiong}}, \ and\ \bibinfo {author}
  {\bibfnamefont {Weisheng}\ \bibnamefont {Zhao}},\ }\bibfield  {title}
  {\enquote {\bibinfo {title} {{Evidence of spin reorientation and
  anharmonicity in kagome ferromagnet Fe$_3$Sn$_2$}},}\ }\href {\doibase
  10.1063/5.0063090} {\bibfield  {journal} {\bibinfo  {journal} {Applied
  Physics Letters}\ }\textbf {\bibinfo {volume} {119}} (\bibinfo {year}
  {2021}),\ 10.1063/5.0063090}\BibitemShut {NoStop}%
\bibitem [{\citenamefont {Xie}\ \emph {et~al.}(2024)\citenamefont {Xie},
  \citenamefont {Deng}, \citenamefont {Zhang}, \citenamefont {Li},
  \citenamefont {Xiong}, \citenamefont {Ma}, \citenamefont {Ma}, \citenamefont
  {Tong}, \citenamefont {Wang}, \citenamefont {Meng}, \citenamefont {Hou},
  \citenamefont {Han}, \citenamefont {Feng},\ and\ \citenamefont
  {Lu}}]{Xie2024}%
  \BibitemOpen
  \bibfield  {author} {\bibinfo {author} {\bibfnamefont {Caihong}\ \bibnamefont
  {Xie}}, \bibinfo {author} {\bibfnamefont {Yongcheng}\ \bibnamefont {Deng}},
  \bibinfo {author} {\bibfnamefont {Dong}\ \bibnamefont {Zhang}}, \bibinfo
  {author} {\bibfnamefont {Junbo}\ \bibnamefont {Li}}, \bibinfo {author}
  {\bibfnamefont {Yimin}\ \bibnamefont {Xiong}}, \bibinfo {author}
  {\bibfnamefont {Mangyuan}\ \bibnamefont {Ma}}, \bibinfo {author}
  {\bibfnamefont {Fusheng}\ \bibnamefont {Ma}}, \bibinfo {author}
  {\bibfnamefont {Wei}\ \bibnamefont {Tong}}, \bibinfo {author} {\bibfnamefont
  {Jihao}\ \bibnamefont {Wang}}, \bibinfo {author} {\bibfnamefont {Wenjie}\
  \bibnamefont {Meng}}, \bibinfo {author} {\bibfnamefont {Yubin}\ \bibnamefont
  {Hou}}, \bibinfo {author} {\bibfnamefont {Yuyan}\ \bibnamefont {Han}},
  \bibinfo {author} {\bibfnamefont {Qiyuan}\ \bibnamefont {Feng}}, \ and\
  \bibinfo {author} {\bibfnamefont {Qingyou}\ \bibnamefont {Lu}},\ }\bibfield
  {title} {\enquote {\bibinfo {title} {{Real‐Space Imaging of Intrinsic
  Symmetry‐Breaking Spin Textures in a Kagome Lattice}},}\ }\href {\doibase
  10.1002/advs.202404088} {\bibfield  {journal} {\bibinfo  {journal} {Advanced
  Science}\ }\textbf {\bibinfo {volume} {11}} (\bibinfo {year} {2024}),\
  10.1002/advs.202404088}\BibitemShut {NoStop}%
\bibitem [{\citenamefont {Wang}\ \emph {et~al.}(2016)\citenamefont {Wang},
  \citenamefont {Sun}, \citenamefont {Zhang}, \citenamefont {Pang},\ and\
  \citenamefont {Lei}}]{Wang2016}%
  \BibitemOpen
  \bibfield  {author} {\bibinfo {author} {\bibfnamefont {Qi}~\bibnamefont
  {Wang}}, \bibinfo {author} {\bibfnamefont {Shanshan}\ \bibnamefont {Sun}},
  \bibinfo {author} {\bibfnamefont {Xiao}\ \bibnamefont {Zhang}}, \bibinfo
  {author} {\bibfnamefont {Fei}\ \bibnamefont {Pang}}, \ and\ \bibinfo {author}
  {\bibfnamefont {Hechang}\ \bibnamefont {Lei}},\ }\bibfield  {title} {\enquote
  {\bibinfo {title} {{Anomalous Hall effect in a ferromagnetic Fe$_3$Sn$_2$
  single crystal with a geometrically frustrated Fe bilayer kagome lattice}},}\
  }\href {\doibase 10.1103/physrevb.94.075135} {\bibfield  {journal} {\bibinfo
  {journal} {Physical Review B}\ }\textbf {\bibinfo {volume} {94}},\ \bibinfo
  {pages} {075135} (\bibinfo {year} {2016})}\BibitemShut {NoStop}%
\bibitem [{\citenamefont {Kumar}\ \emph {et~al.}(2019)\citenamefont {Kumar},
  \citenamefont {Soh}, \citenamefont {Wang},\ and\ \citenamefont
  {Xiong}}]{Kumar2019}%
  \BibitemOpen
  \bibfield  {author} {\bibinfo {author} {\bibfnamefont {Neeraj}\ \bibnamefont
  {Kumar}}, \bibinfo {author} {\bibfnamefont {Y.}~\bibnamefont {Soh}}, \bibinfo
  {author} {\bibfnamefont {Yihao}\ \bibnamefont {Wang}}, \ and\ \bibinfo
  {author} {\bibfnamefont {Y.}~\bibnamefont {Xiong}},\ }\bibfield  {title}
  {\enquote {\bibinfo {title} {{Magnetotransport as a diagnostic of spin
  reorientation: Kagome ferromagnet as a case study}},}\ }\href {\doibase
  10.1103/physrevb.100.214420} {\bibfield  {journal} {\bibinfo  {journal}
  {Physical Review B}\ }\textbf {\bibinfo {volume} {100}},\ \bibinfo {pages}
  {214420} (\bibinfo {year} {2019})}\BibitemShut {NoStop}%
\bibitem [{\citenamefont {Fitch}\ \emph {et~al.}(2023)\citenamefont {Fitch},
  \citenamefont {Dejoie}, \citenamefont {Covacci}, \citenamefont
  {Confalonieri}, \citenamefont {Grendal}, \citenamefont {Claustre},
  \citenamefont {Guillou}, \citenamefont {Kieffer}, \citenamefont {de~Nolf},
  \citenamefont {Petitdemange}, \citenamefont {Ruat},\ and\ \citenamefont
  {Watier}}]{Fitch2023}%
  \BibitemOpen
  \bibfield  {author} {\bibinfo {author} {\bibfnamefont {Andrew}\ \bibnamefont
  {Fitch}}, \bibinfo {author} {\bibfnamefont {Catherine}\ \bibnamefont
  {Dejoie}}, \bibinfo {author} {\bibfnamefont {Ezio}\ \bibnamefont {Covacci}},
  \bibinfo {author} {\bibfnamefont {Giorgia}\ \bibnamefont {Confalonieri}},
  \bibinfo {author} {\bibfnamefont {Ola}\ \bibnamefont {Grendal}}, \bibinfo
  {author} {\bibfnamefont {Laurent}\ \bibnamefont {Claustre}}, \bibinfo
  {author} {\bibfnamefont {Perceval}\ \bibnamefont {Guillou}}, \bibinfo
  {author} {\bibfnamefont {Jérôme}\ \bibnamefont {Kieffer}}, \bibinfo
  {author} {\bibfnamefont {Wout}\ \bibnamefont {de~Nolf}}, \bibinfo {author}
  {\bibfnamefont {Sébastien}\ \bibnamefont {Petitdemange}}, \bibinfo {author}
  {\bibfnamefont {Marie}\ \bibnamefont {Ruat}}, \ and\ \bibinfo {author}
  {\bibfnamefont {Yves}\ \bibnamefont {Watier}},\ }\bibfield  {title} {\enquote
  {\bibinfo {title} {{ID22 – the high-resolution powder-diffraction beamline
  at ESRF}},}\ }\href {\doibase 10.1107/s1600577523004915} {\bibfield
  {journal} {\bibinfo  {journal} {Journal of Synchrotron Radiation}\ }\textbf
  {\bibinfo {volume} {30}},\ \bibinfo {pages} {1003--1012} (\bibinfo {year}
  {2023})}\BibitemShut {NoStop}%
\bibitem [{\citenamefont {Tsirlin}\ \emph {et~al.}(2022)\citenamefont
  {Tsirlin}, \citenamefont {Prodan},\ and\ \citenamefont
  {Grendal}}]{TSIRLIN2025}%
  \BibitemOpen
  \bibfield  {author} {\bibinfo {author} {\bibfnamefont {Alexander~A.}\
  \bibnamefont {Tsirlin}}, \bibinfo {author} {\bibfnamefont {Lilian}\
  \bibnamefont {Prodan}}, \ and\ \bibinfo {author} {\bibfnamefont {Ola}\
  \bibnamefont {Grendal}},\ }\bibfield  {title} {\enquote {\bibinfo {title}
  {{High-resolution x-ray diffraction study of the kagome metal
  Fe$_3$Sn$_2$}},}\ }\href {\doibase 10.15151/ESRF-DC-2007330151} {\bibfield
  {journal} {\bibinfo  {journal} {European Synchrotron Radiation Facility}\ }
  (\bibinfo {year} {2022}),\ 10.15151/ESRF-DC-2007330151}\BibitemShut {NoStop}%
\bibitem [{\citenamefont {Pet{\u r}{\'\i}{\v c}ek}\ \emph
  {et~al.}(2014)\citenamefont {Pet{\u r}{\'\i}{\v c}ek}, \citenamefont {Du{\v
  s}ek},\ and\ \citenamefont {Palatinus}}]{jana2006}%
  \BibitemOpen
  \bibfield  {author} {\bibinfo {author} {\bibfnamefont {V.}~\bibnamefont
  {Pet{\u r}{\'\i}{\v c}ek}}, \bibinfo {author} {\bibfnamefont
  {M.}~\bibnamefont {Du{\v s}ek}}, \ and\ \bibinfo {author} {\bibfnamefont
  {L.}~\bibnamefont {Palatinus}},\ }\bibfield  {title} {\enquote {\bibinfo
  {title} {Crystallographic computing system {JANA2006}: General features},}\
  }\href {\doibase 10.1515/zkri-2014-1737} {\bibfield  {journal} {\bibinfo
  {journal} {Z. Krist.}\ }\textbf {\bibinfo {volume} {229}},\ \bibinfo {pages}
  {345--352} (\bibinfo {year} {2014})}\BibitemShut {NoStop}%
\bibitem [{\citenamefont {Stuhr}\ \emph {et~al.}(2017)\citenamefont {Stuhr},
  \citenamefont {Roessli}, \citenamefont {Gvasaliya}, \citenamefont {Rønnow},
  \citenamefont {Filges}, \citenamefont {Graf}, \citenamefont {Bollhalder},
  \citenamefont {Hohl}, \citenamefont {Bürge}, \citenamefont {Schild},
  \citenamefont {Holitzner}, \citenamefont {Kaegi}, \citenamefont {Keller},\
  and\ \citenamefont {Mühlebach}}]{Stuhr2017}%
  \BibitemOpen
  \bibfield  {author} {\bibinfo {author} {\bibfnamefont {U.}~\bibnamefont
  {Stuhr}}, \bibinfo {author} {\bibfnamefont {B.}~\bibnamefont {Roessli}},
  \bibinfo {author} {\bibfnamefont {S.}~\bibnamefont {Gvasaliya}}, \bibinfo
  {author} {\bibfnamefont {H.M.}\ \bibnamefont {Rønnow}}, \bibinfo {author}
  {\bibfnamefont {U.}~\bibnamefont {Filges}}, \bibinfo {author} {\bibfnamefont
  {D.}~\bibnamefont {Graf}}, \bibinfo {author} {\bibfnamefont {A.}~\bibnamefont
  {Bollhalder}}, \bibinfo {author} {\bibfnamefont {D.}~\bibnamefont {Hohl}},
  \bibinfo {author} {\bibfnamefont {R.}~\bibnamefont {Bürge}}, \bibinfo
  {author} {\bibfnamefont {M.}~\bibnamefont {Schild}}, \bibinfo {author}
  {\bibfnamefont {L.}~\bibnamefont {Holitzner}}, \bibinfo {author}
  {\bibfnamefont {C.}~\bibnamefont {Kaegi}}, \bibinfo {author} {\bibfnamefont
  {P.}~\bibnamefont {Keller}}, \ and\ \bibinfo {author} {\bibfnamefont
  {T.}~\bibnamefont {Mühlebach}},\ }\bibfield  {title} {\enquote {\bibinfo
  {title} {{The thermal triple-axis-spectrometer EIGER at the continuous
  spallation source SINQ}},}\ }\href {\doibase 10.1016/j.nima.2017.02.003}
  {\bibfield  {journal} {\bibinfo  {journal} {Nuclear Instruments and Methods
  in Physics Research Section A: Accelerators, Spectrometers, Detectors and
  Associated Equipment}\ }\textbf {\bibinfo {volume} {853}},\ \bibinfo {pages}
  {16--19} (\bibinfo {year} {2017})}\BibitemShut {NoStop}%
\bibitem [{\citenamefont {Kazakova}\ \emph {et~al.}(2019)\citenamefont
  {Kazakova}, \citenamefont {Puttock}, \citenamefont {Barton}, \citenamefont
  {Corte-León}, \citenamefont {Jaafar}, \citenamefont {Neu},\ and\
  \citenamefont {Asenjo}}]{Kazakova2019}%
  \BibitemOpen
  \bibfield  {author} {\bibinfo {author} {\bibfnamefont {O.}~\bibnamefont
  {Kazakova}}, \bibinfo {author} {\bibfnamefont {R.}~\bibnamefont {Puttock}},
  \bibinfo {author} {\bibfnamefont {C.}~\bibnamefont {Barton}}, \bibinfo
  {author} {\bibfnamefont {H.}~\bibnamefont {Corte-León}}, \bibinfo {author}
  {\bibfnamefont {M.}~\bibnamefont {Jaafar}}, \bibinfo {author} {\bibfnamefont
  {V.}~\bibnamefont {Neu}}, \ and\ \bibinfo {author} {\bibfnamefont
  {A.}~\bibnamefont {Asenjo}},\ }\bibfield  {title} {\enquote {\bibinfo {title}
  {{Frontiers of magnetic force microscopy}},}\ }\href {\doibase
  10.1063/1.5050712} {\bibfield  {journal} {\bibinfo  {journal} {Journal of
  Applied Physics}\ }\textbf {\bibinfo {volume} {125}} (\bibinfo {year}
  {2019}),\ 10.1063/1.5050712}\BibitemShut {NoStop}%
\bibitem [{\citenamefont {O'Handley}(1999)}]{Ohandley1999modern}%
  \BibitemOpen
  \bibfield  {author} {\bibinfo {author} {\bibfnamefont {Robert~C.}\
  \bibnamefont {O'Handley}},\ }\href@noop {} {\emph {\bibinfo {title} {Modern
  Magnetic Materials: Principles and Applications}}}\ (\bibinfo  {publisher}
  {Wiley},\ \bibinfo {address} {New York},\ \bibinfo {year} {1999})\ p.\
  \bibinfo {pages} {184}\BibitemShut {NoStop}%
\bibitem [{\citenamefont {Cullity}\ and\ \citenamefont
  {Graham}(2008)}]{Cullity2008}%
  \BibitemOpen
  \bibfield  {author} {\bibinfo {author} {\bibfnamefont {B.~D.}\ \bibnamefont
  {Cullity}}\ and\ \bibinfo {author} {\bibfnamefont {C.~D.}\ \bibnamefont
  {Graham}},\ }\href {\doibase 10.1002/9780470386323} {\emph {\bibinfo {title}
  {Introduction to Magnetic Materials}}}\ (\bibinfo  {publisher} {Wiley},\
  \bibinfo {year} {2008})\BibitemShut {NoStop}%
\bibitem [{\citenamefont {Prodan}\ \emph {et~al.}(2023)\citenamefont {Prodan},
  \citenamefont {Evans}, \citenamefont {Griffin}, \citenamefont {Östlin},
  \citenamefont {Altthaler}, \citenamefont {Lysne}, \citenamefont {Filippova},
  \citenamefont {Shova}, \citenamefont {Chioncel}, \citenamefont {Tsurkan},\
  and\ \citenamefont {Kézsmárki}}]{Prodan2023}%
  \BibitemOpen
  \bibfield  {author} {\bibinfo {author} {\bibfnamefont {Lilian}\ \bibnamefont
  {Prodan}}, \bibinfo {author} {\bibfnamefont {Donald~M.}\ \bibnamefont
  {Evans}}, \bibinfo {author} {\bibfnamefont {Sinéad~M.}\ \bibnamefont
  {Griffin}}, \bibinfo {author} {\bibfnamefont {Andreas}\ \bibnamefont
  {Östlin}}, \bibinfo {author} {\bibfnamefont {Markus}\ \bibnamefont
  {Altthaler}}, \bibinfo {author} {\bibfnamefont {Erik}\ \bibnamefont {Lysne}},
  \bibinfo {author} {\bibfnamefont {Irina~G.}\ \bibnamefont {Filippova}},
  \bibinfo {author} {\bibfnamefont {Sergiu}\ \bibnamefont {Shova}}, \bibinfo
  {author} {\bibfnamefont {Liviu}\ \bibnamefont {Chioncel}}, \bibinfo {author}
  {\bibfnamefont {Vladimir}\ \bibnamefont {Tsurkan}}, \ and\ \bibinfo {author}
  {\bibfnamefont {István}\ \bibnamefont {Kézsmárki}},\ }\bibfield  {title}
  {\enquote {\bibinfo {title} {{Large ordered moment with strong easy-plane
  anisotropy and vortex-domain pattern in the kagome ferromagnet Fe$_3$Sn}},}\
  }\href {\doibase 10.1063/5.0155295} {\bibfield  {journal} {\bibinfo
  {journal} {Applied Physics Letters}\ }\textbf {\bibinfo {volume} {123}}
  (\bibinfo {year} {2023}),\ 10.1063/5.0155295}\BibitemShut {NoStop}%
\bibitem [{\citenamefont {Sucksmith}\ and\ \citenamefont
  {Thompson}(1954)}]{Sucksmith1954}%
  \BibitemOpen
  \bibfield  {author} {\bibinfo {author} {\bibfnamefont {Willie}\ \bibnamefont
  {Sucksmith}}\ and\ \bibinfo {author} {\bibfnamefont {Jo~E}\ \bibnamefont
  {Thompson}},\ }\bibfield  {title} {\enquote {\bibinfo {title} {The magnetic
  anisotropy of cobalt},}\ }\href {\doibase 10.1098/rspa.1954.0209} {\bibfield
  {journal} {\bibinfo  {journal} {Proceedings of the Royal Society of London.
  Series A. Mathematical and Physical Sciences}\ }\textbf {\bibinfo {volume}
  {225}},\ \bibinfo {pages} {362--375} (\bibinfo {year} {1954})}\BibitemShut
  {NoStop}%
\bibitem [{\citenamefont {Hubert}\ and\ \citenamefont
  {Schäfer}(2009)}]{Hubert2009}%
  \BibitemOpen
  \bibfield  {author} {\bibinfo {author} {\bibfnamefont {A.}~\bibnamefont
  {Hubert}}\ and\ \bibinfo {author} {\bibfnamefont {R.}~\bibnamefont
  {Schäfer}},\ }\href@noop {} {\emph {\bibinfo {title} {{Magnetic Domains: The
  Analysis of Magnetic Microstructures}}}},\ Corrected 3rd Printing\ (\bibinfo
  {publisher} {Springer},\ \bibinfo {address} {Berlin, Heidelberg, New York},\
  \bibinfo {year} {2009})\ p.\ \bibinfo {pages} {378}\BibitemShut {NoStop}%
\bibitem [{\citenamefont {Chen}\ \emph {et~al.}(2021)\citenamefont {Chen},
  \citenamefont {Lv}, \citenamefont {Wu}, \citenamefont {Hu}, \citenamefont
  {Li}, \citenamefont {Wang}, \citenamefont {Xiong}, \citenamefont {Gao},
  \citenamefont {Tang}, \citenamefont {Tian},\ and\ \citenamefont
  {Du}}]{Chen2021}%
  \BibitemOpen
  \bibfield  {author} {\bibinfo {author} {\bibfnamefont {Yutao}\ \bibnamefont
  {Chen}}, \bibinfo {author} {\bibfnamefont {Boyao}\ \bibnamefont {Lv}},
  \bibinfo {author} {\bibfnamefont {Yaodong}\ \bibnamefont {Wu}}, \bibinfo
  {author} {\bibfnamefont {Qiyang}\ \bibnamefont {Hu}}, \bibinfo {author}
  {\bibfnamefont {Junbo}\ \bibnamefont {Li}}, \bibinfo {author} {\bibfnamefont
  {Yihao}\ \bibnamefont {Wang}}, \bibinfo {author} {\bibfnamefont {Yimin}\
  \bibnamefont {Xiong}}, \bibinfo {author} {\bibfnamefont {Jianhua}\
  \bibnamefont {Gao}}, \bibinfo {author} {\bibfnamefont {Jin}\ \bibnamefont
  {Tang}}, \bibinfo {author} {\bibfnamefont {Mingliang}\ \bibnamefont {Tian}},
  \ and\ \bibinfo {author} {\bibfnamefont {Haifeng}\ \bibnamefont {Du}},\
  }\bibfield  {title} {\enquote {\bibinfo {title} {{Effects of tilted
  magnetocrystalline anisotropy on magnetic domains in Fe$_3$Sn$_2$ thin
  plates}},}\ }\href {\doibase 10.1103/physrevb.103.214435} {\bibfield
  {journal} {\bibinfo  {journal} {Physical Review B}\ }\textbf {\bibinfo
  {volume} {103}},\ \bibinfo {pages} {214435} (\bibinfo {year}
  {2021})}\BibitemShut {NoStop}%
\end{thebibliography}%
